# Development of a Deep Reinforcement Learning Approach for Transition State Search in Molecular Systems: A Transferable Model Across Structural and Dimensional Variations

Takechika Kikkawa[1], Shusuke Yamanaka[1,2], Takashi Kawakami[1], Mitsutaka Okumura[1,3]

1. Graduate School of Science, Osaka University, 1-1, Machikaneyama, Toyonaka, Osaka 560-0043, Japan.
2. Center of Quantum Information and Quantum Biology, Osaka University, 1-2, Machikaneyama, Toyonaka, Osaka, 560-0043, Japan.
3. Innovative Catalysis Science Division, Institute for Open and Transdisciplinary Research Initiatives, 1-1 Yamada, Suita, Osaka, 656-0871, Japan.

**Abstract**

We developed a DRL approach for TS searches combining the deep Q-network and the bidirectional long short-term memory architectures to handle variable-sized molecular systems. We applied our method to 1,3,5-hexatriene (Triene), alanine dipeptide (Ala2), and 1,3,5,7-octatetraene (Tetraene) as target systems with the search space limited to backbone dihedral angles. Our model successfully learned the TS search tasks for all examined systems. More importantly, we demonstrate that our model pre-trained on Triene exhibits distinct transfer-learning capabilities across both backbone modifications (Triene to Ala2) and chain extensions (Triene to Tetraene). Such transferability is highly promising for future applications to molecules in general, as it allows an agent trained on simple systems to accelerate TS searches in larger or more complex ones. We expect that these findings on cross-system transferability represent an important step forward in developing DRL-based approaches for TS searches.

## 1. Introduction

Finding transition state (TS) structures on potential energy surfaces (PESs) is essential for understanding chemical reaction mechanisms. Once a TS structure is located, the chemical reaction path can be identified as the intrinsic reaction coordinate (IRC) [1], which connects the TS to the nearest equilibrium (EQ) structures. The problem is that, because TS structures correspond to first-order saddle points on PESs, there is no mathematically established procedures guaranteed to find them, unlike the minimization problem. Thus, computational chemists have devoted significant effort to developing TS search methods [2-24]. While these include double-ended methods that require structural information of both reactants and products [2-8], our focus is on single-ended methods that start from a reactant structure, i.e., a single stable molecular geometry, and navigate uphill to a TS structure on the PES. One of the most important single-ended approaches is the eigenvector-following (EVF) method [11-15]. This method is designed to climb from an EQ state to a TS by maximizing the energy along one direction, specifically the direction of the smallest curvature, while simultaneously minimizing the energy along the other 3N-7 directions for a molecular system consisting of N-atoms. A more sophisticated approach is the anharmonic downward distortion following (ADDF) method developed by Maeda and Ohno, which climbs from an EQ structure to the TS structures by sequentially tracking the minimum points on the iso-energy hypersurface constructed in scaled normal coordinates [16-18]. Like the EVF and ADDF methods, most other effective methods share the common feature of utilizing not only the gradients but also the curvatures of the PES to detect changes in curvature [2,3]. In contrast, some researchers have proposed methods that utilize only the gradient [19-24]. For instance, the dimer method locates the TS by following the lowest curvature path on the PES, detected by a rotating dimer [19,20]. Similarly, the gradient-only algorithm developed by Quapp is designed to trace the valley-floor route in the uphill direction, along which $\nabla \mathrm{E}(\boldsymbol{x} + \nabla \mathrm{E}(\boldsymbol{x})) \propto \nabla \mathrm{E}(\boldsymbol{x})$ holds [21]. The global area search (GLAS) method recently proposed by Shoji et al. applies repulsive interactions among several molecular structures termed "walkers" to drive them uphill along paths of relatively small curvature [24,25]. Although these approaches use only gradients, they effectively capture the direction of lowest curvature. We would like to emphasize that, in all the methods described here, the algorithms are essentially designed to follow the plausible path along which one and only one curvature changes from positive (EQ) to negative (TS).

Instead of the mechanics-driven approaches listed above, many researchers have recently developed data-driven approaches—specifically, various supervised machine learning (ML) architectures—for the TS search task [26–33]. For instance, Pattanaik et al. proposed a graph neural network (NN) model to predict the TS distance matrix [27]. Similarly, Jackson et al. [28] and Choi [29] developed a message-passing NN and a transformer-GRU-based model, respectively, to generate TS geometries. Advanced generative models, including diffusion and flow architectures, were also introduced by Duan et al. [30], Kim et al. [31], and Luo et al. [32]. However, all of these architectures require structural information of both reactants and products. In contrast, Hayashi et al. pioneered a single-ended generative architecture, which requires only reactant structures and reaction-type labels as input to generate reaction paths [33]. Subsequently, Galustian et al. formulated TS generation as an optimal transport flow problem [34], and more recently, Xu et al. presented a diffusion model to generate 3D TS structures from 2D reactant graphs, alongside a dataset comprising 4,391 TS structures [35]. While all of these architectures represent state-of-the-art approaches and innovative solutions to the TS search problem, such supervised ML approaches inherently rely on the quantity and quality of training data. In practice, available training datasets for TS structures are still far from comprehensive across the vast, chemically relevant space, thereby limiting the broader applicability of ML-based TS search methods.

To overcome this limitation, reinforcement learning (RL), a paradigm distinct from supervised learning, has been increasingly developed for searching chemical reaction pathways [36–42], as it inherently does not require pre-existing training data. For alloy systems, RL has been utilized to model transition kinetics, such as predicting surface reconstruction pathways [35] and simulating hydrogen transport alongside an ML potential [36]. In heterogeneous catalysis, Lan and An coupled deep RL with DFT calculations to search across discrete surface-adsorption states rather than continuous atomic geometries, uncovering a novel multistep Haber–Bosch pathway [37]. Zhang et al. took a dynamical approach to chemical reactions and formulated the TS search as a shooting game, in which an agent launches two trajectories with opposite momenta from a candidate structure and succeeds if both trajectories reach the reactant and product states [38]. Closer to conventional quantum chemical TS search, Xu et al. applied Q-learning to locate TS structures for two simple gas-phase reactions [39]; however, this method inherently requires fully pre-calculated PESs over the reaction coordinates. Similarly, relying on the prior knowledge of large-scale, pre-computed reaction path networks, Matsumura et al. benchmarked several reinforcement learning algorithms for connecting reactant and product states [40]. Several studies have also reported RL-based quantum chemical TS searches in a double-ended framework without prior knowledge of the global potential energy surface (PES) [41, 42]—an essential requirement for practical TS search approaches. Barrett and Westermayr combined an actor-critic RL model with the nudged elastic band (NEB) method, successfully applying it to a Claisen rearrangement and several $S_N2$ reactions [41]. Concurrently, Suresh and Vogiatzis developed a soft actor-critic framework to locate TS structures across eight simple molecular reactions [42].

In contrast to the RL approaches cited above, we previously proposed a straightforward RL model to learn efficient pathways from an EQ state to the TS region on PESs through trial and error [43]—that is, without introducing elaborate setups or requiring a priori information about the product state or the global PES. This represents the most natural formulation of RL for the TS search task, given that RL is inherently designed for such spatial navigation problems; finding a goal in a maze, for instance, is one of the fundamental benchmark tasks in RL [44,45]. Along these lines, RL approaches for locating energy minima [46] or improving geometry optimization efficiency [47] have also been developed. Despite the success of RL in geometry optimization tasks, its application to TS search remains limited due to the inherent difficulty of formulating an appropriate model to find saddle points on high-dimensional energy surfaces. To address this challenge, we integrated the established knowledge of conventional quantum-chemical TS search algorithms discussed above into an RL approach employing a deep Q-network (DQN)—specifically, by using PES gradients, curvatures, and changes in curvature as the state representation [43]. This distinguishes our approach from the prior DRL study by Pal [48], which employed only positions in coordinate space as the state. Indeed, we found that even relying solely on gradients in our DQN model [45,49] resulted in poor learning, whereas adding these curvature features to the state led to a substantial improvement in learning performance (see the Supporting Information of Ref. [43]). Furthermore, we demonstrated that the RL agent could leverage its prior TS search experience from several PESs to accelerate the learning process on unknown PESs [43]. This transferability implies that our model learned the underlying principles of locating TS structures using local PES features—effectively acquiring the strategies established over decades by computational chemists. However, the targets in our previous study were the two-dimensional (2D) model PESs, rather than molecular systems.

In this study, we apply our DQN model to several molecular systems. To the best of our knowledge, this is the first application of a DRL approach to single-ended TS searches in molecular systems without requiring prior knowledge of PESs. The application of this method to realistic targets, however, is nontrivial. The first hurdle is the numerous trial-and-error DRL runs needed for the learning process, making exploration over the full 3N-6 dimensions of realistic $N$-atom molecules computationally demanding. Thus, to reduce the dimensionality for the DRL runs, we selected simple molecular systems for benchmark calculations, and restricted the search spaces to the backbone dihedral spaces. The second problem lies in generalizing a single DRL model across molecules of different sizes, which is essential for establishing this DRL-based framework as a universal TS search method for molecular systems. To address this issue, we integrated a bidirectional long short-term memory (BiLSTM) network [50] into our DQN model, enabling the transfer of learned TS search capabilities from specific molecules to others of different sizes. To validate this extended model, we performed benchmark calculations using 1,3,5-hexatriene (Triene), alanine dipeptide (Ala2), and 1,3,5,7-octatetraene (Tetraene) to examine whether our model trained on Triene exhibits transferability under backbone modification (Triene to Ala2) and chain extension (Triene to Tetraene). The computational results successfully demonstrated that our agent possesses the desired transferability across these structural and dimensional variations. Based on contemporary transfer-learning theory, we argue that integrating the key principles of quantum-chemical TS search methods provides a crucial inductive bias, enabling these transfer effects. Finally, we provide a perspective on the potential advantages of our model's transferability regarding its practical implementation in molecular systems.

## 2. Theoretical Background

### 2-1. Deep Q Learning algorithm [45,49]

In this section, we briefly describe the deep reinforcement learning (DRL) techniques employed in our research. We model the TS search task as a Markov decision process (MDP) characterized by the transition probability $p(s_{t+1}, r_{t+1}|s_t, a_t)$, which provides the probability that the agent taking an action $a_t$ in a state $s_t$ at a time $t$ moves to a new state $s_{t+1}$ and receive a reward $r_{t+1}$. The learning process is divided into subsequences termed "episodes" such as individual games, maze navigations, or driving trials [44,45]. During each episode, the agent learns an optimal policy to maximize the expected discounted return up to the termination step, $T$,

$$G_t = r_{t+1} + \gamma r_{t+2} + \cdots + \gamma^{T-t-1} r_T. \quad (1)$$

In the Q-learning algorithm that serves as RL framework of our method, the agent's performance is evaluated on the basis of the action-value function,

$$Q(\boldsymbol{s}, \boldsymbol{a}) = \mathbb{E}[G_t|s_t = \boldsymbol{s}, a_t = \boldsymbol{a}], \quad (2)$$

where $\mathbb{E}$ denotes the expectation over the transition probability of the MDP, $p(s_{t+1}, r_{t+1}|s_t, a_t)$, starting from a state $\boldsymbol{s}$, and taking an action, $\boldsymbol{a}$. To obtain the optimal action-value function under conditions where the explicit form of the transition probability, $p(s_{t+1}, r_{t+1}|s_t, a_t)$ is unknown, the Q-function is iteratively updated with a learning rate $\alpha$ $(0 < \alpha \leq 1)$ following each action of the agent, as given by [44]

$$Q(s_t, a_t) \leftarrow Q(s_t, a_t) + \alpha\big(r_{t+1} + \gamma Q(s_{t+1}, a_{t+1}) - Q(s_t, a_t)\big). \quad (3)$$

In the traditional Q-learning approach, a Q-table is maintained for state–action pairs, which becomes computationally infeasible when the state-action space is high-dimensional. To address this problem, we employ the deep Q-Networks (DQN) architecture [45,49], in which the Q-function, $Q(\boldsymbol{s}, \boldsymbol{a})$, is approximated by a deep neural network (DNN) called "Q-Network", $Q_\theta(\boldsymbol{s}, \boldsymbol{a})$. While this DNN is typically a multilayer perceptron (MLP), it can encompass a wide variety of other architectures, such as convolutional or recurrent neural networks [45]. The specific network used in this study is detailed in Section 2.4. Regardless of the architecture, rather than the Q-function itself, the parameter set $\theta$ of the DNN is updated to minimize the loss function,

$$\theta \leftarrow \text{argmin}_\theta\, \mathcal{L}[r_t + \gamma\, \text{Max}_{a'}\, Q_\theta(s_{t+1}, a') - Q_\theta(s_t, a_t)]). \quad (4)$$

Following the original paper of DQN [49], we employ the Huber loss function for, $\mathcal{L}(\theta)$. We also employ two other techniques from the original DQN: the first is an experience replay scheme to mitigate the strong correlations inevitable in sequential transition data. This method stores the experiences $\{s_t, a_t, s_{t+1}, r_{t+1}\}$ into a buffer, and then samples them randomly from the buffer, rather than from consecutive time steps, to update the Q-network, $Q_\theta$. The second is the use of a target network $Q_\theta^{target}$, which is a separate copy of the Q-network. While the parameter set $\theta$ of the primary Q-network $Q_\theta$ is updated at every step, $Q_\theta^{target}$ is synchronized with $Q_\theta$ every C steps, where the constant C is set to 1000 steps in our application. The parameter set $\theta$ is randomly initialized at the outset of the learning process and is carried over to the subsequent episodes. In order to avoid the dependence on the initial parameters, all results reported throughout this study are averaged over 10 independent runs initialized with different random sets of $\theta$.

### 2-2. Environment design [43]

In this subsection, we review the environment design for our DRL model adopted in the present study, which was successfully applied to two-dimensional systems in our previous work [43]. The objective of our method is to locate a transition state (TS) structure $\boldsymbol{x}_{\mathrm{TS}}$ starting from an equilibrium (EQ) structure $\boldsymbol{x}_{\mathrm{EQ}}$ on a potential energy surface (PES), $\mathrm{E}(\boldsymbol{x})$. These structures correspond to stationary points on the PES, where the gradient ($\boldsymbol{g}(\boldsymbol{x}) \equiv \nabla \mathrm{E}(\boldsymbol{x})$) vanishes:

$$\boldsymbol{g}(\boldsymbol{x}_{\mathrm{EQ}}) = \boldsymbol{g}(\boldsymbol{x}_{\mathrm{TS}}) = \mathbf{0}. \qquad (5)$$

The distinction between TS and EQ lies in the curvatures defined as the eigenvalues of the mass-weighted Hessian matrix $\mathbf{H}(\boldsymbol{x})$ for a molecular system consisting of N atoms:

$$\mathbf{H}(\boldsymbol{x})\boldsymbol{v}_i(\boldsymbol{x}) = \omega_i \boldsymbol{v}_i(\boldsymbol{x}),\ i = 1\sim 3\mathrm{N}-6 \quad (6)$$

The EQ structure is a local minimum, where all of the curvature values are positive, i.e.,

$$0 < \omega_1(\boldsymbol{x}_{\mathrm{EQ}}) \leq \cdots \leq \omega_{3\mathrm{N}-6}(\boldsymbol{x}_{\mathrm{EQ}}) \qquad (7)$$

while the TS structure is a saddle point characterized by one negative curvature and positive curvatures for all other modes, i.e.,

$$\omega_1(\boldsymbol{x}_{\mathrm{TS}}) < 0 < \omega_2(\boldsymbol{x}_{\mathrm{TS}}) \leq \cdots \leq \omega_{3\mathrm{N}-6}(\boldsymbol{x}_{\mathrm{TS}}) \qquad (8)$$

To reach the TS region, the RL agent must first move out of the basin of an EQ point where all curvatures are positive, and walk uphill along a direction in which one of the curvatures changes from positive to negative. Thus, in order to detect changes in curvatures, the agent receives from the environment not only the first- and second-order derivatives (gradients and curvatures), but also the third-order derivatives (changes in curvatures) of the potential energy function as the state information $s_t$. This environment setup is inspired by the established computational chemistry methods for identifying the chemical reaction paths as described in the Introduction. Accordingly, we define the state components based on up to the third-order derivatives as follows:

$$\boldsymbol{f}(\boldsymbol{x}_t) = \left(f_1(\boldsymbol{x}_t), \ldots, f_{3\mathrm{N}-6}(\boldsymbol{x}_t)\right) = \left(\tanh\left(-\frac{g_1(\boldsymbol{x}_t)}{\omega_1(\boldsymbol{x}_t)}\right), \ldots, \tanh\left(-\frac{g_{3\mathrm{N}-6}(\boldsymbol{x}_t)}{\omega_{3\mathrm{N}-6}(\boldsymbol{x}_t)}\right)\right), \quad (9)$$

$$\boldsymbol{c}(\boldsymbol{x}_t) = \left(c_1(\boldsymbol{x}_t), \ldots, c_{3\mathrm{N}-6}(\boldsymbol{x}_t)\right) = \left(\frac{\min_i(|\omega_i(\boldsymbol{x}_t)|)}{\omega_1(\boldsymbol{x}_t)}, \ldots, \frac{\min_i(|\omega_i(\boldsymbol{x}_t)|)}{\omega_{3\mathrm{N}-6}(\boldsymbol{x}_t)}\right), \quad (10)$$

$$\boldsymbol{q}(\boldsymbol{x}_t) = \left(q_1(\boldsymbol{x}_t), \dots, q_{3\mathrm{N}-6}(\boldsymbol{x}_t)\right) = \left(\frac{\Delta\omega_1(x_t)}{|\omega_1(x_t)|}, \dots, \frac{\Delta\omega_{3\mathrm{N}-6}(x_t)}{|\omega_{3\mathrm{N}-6}(x_t)|}\right). \qquad (11)$$

Here, to stabilize the learning process, these variables are transformed as defined in Eqs. (9)-(11). For the first-order information given by eq.(9), rather than using gradients directly, we employ a vector consisting of tanh transformations of the Newton-Raphson parameters. In eq. (10), the second-order derivative information consist of the inverse curvatures, scaled by their minimum absolute value. All elements of these two vectors are constrained to the range [-1,1]. Each element of the third-order derivative information given by eq. (11) is divided by the absolute value of the corresponding curvature element. Furthermore, to prevent mismatches of eigenvectors between successive steps when estimating the changes of curvatures, $\Delta\omega_i(\boldsymbol{x}_t) \equiv \omega_i(\boldsymbol{x}_t) - \omega_i(\boldsymbol{x}_{t-1})$, in eq. (11), we sort the sets of eigenvectors and eigenvalues at each step. Moreover, all elements of $\boldsymbol{f}, \boldsymbol{c},$ and $\boldsymbol{q},$ remain invariant to the scaling of the PES curvatures, thereby ensuring the transferability of the Q-table domain across PESs with different curvature magnitudes. These vectors are concatenated into a single vector,

$$\boldsymbol{d}(\boldsymbol{x}_t) = (d_1(\boldsymbol{x}_t), \dots, d_{3N-6}(\boldsymbol{x}_t)), \qquad (12)$$

$$d_i(\boldsymbol{x}_t) = (f_i(\boldsymbol{x}_t), c_i(\boldsymbol{x}_t), q_i(\boldsymbol{x}_t)), \;\; 1 \le i \le 3\mathrm{N}-6, \qquad (13)$$

and, following the study by Ahuja et al., the state representation is defined by stacking the $\boldsymbol{d}$ vectors from the current and four preceding time steps:

$$\boldsymbol{s}(\boldsymbol{x}_t) = \left(s_1(\boldsymbol{x}_t), \dots, s_{3N-6}(\boldsymbol{x}_t)\right), \qquad (14)$$

$$s_i(\boldsymbol{x}_t) \equiv (d_i(\boldsymbol{x}_{t-4}), d_i(\boldsymbol{x}_{t-3}), d_i(\boldsymbol{x}_{t-2}), d_i(\boldsymbol{x}_{t-1}), d_i(\boldsymbol{x}_t)), \;\; 1 \le i \le 3\mathrm{N}-6 \qquad (15)$$

which is intended to capture the local neighborhood information around $\boldsymbol{x}_t$.

### 2.3. Action and Reward Design

To select the action, $\boldsymbol{a}_t$ at $\boldsymbol{s}(\boldsymbol{x}_t)$ , we employ a time-dependent greedy method as follows. The key problem is how to balance the exploration and exploitation of actions. The RL agent should take the greedy action that yields the maximal Q-value in almost all cases after the Q-function has been sufficiently trained. In contrast, because the Q -network is unreliable during the early stages of training from scratch, it is preferable to select random actions as well. This allows the agent to discover superior actions that the untrained Q-network might otherwise overlook during greedy selection. Thus, in our model, the agent selects an action uniformly at random with probability:

$$\epsilon(t) = \mathrm{Max}\,(0.5 - 0.4\,\mathrm{N}_{sat}^{-1}\,t, 0.1), \qquad (16)$$

and the greedy action at the rate of $1 - \epsilon(t)$. Here, $t$ denotes the total number of the training steps taken by the agent, which we expect to serve as an indicator of its training progress. Note that although the initial exploration rate is set to $\epsilon(0) = 0.5$, the agent's action selection is effectively completely random at the onset of training because the parameter set $\theta$ of the Q-table is randomly initialized. This setting allows the agent quickly shift toward optimal actions if the optimization of $\theta$ proceeds rapidly, since the exploration rate is linearly decay from 0.5 , rather than the conventional 1.0, to a final value, 0.1, at $\mathrm{N}_{sat}$. We set $\mathrm{N}_{sat} = 2.0 \times 10^5$ in all the following calculations.

The possible actions of the agent are to move forward (+) or backward (-) along one of the 3N-6 directions, $\{\boldsymbol{v}_1(\boldsymbol{x}_t), \dots, \boldsymbol{v}_{3\mathrm{N}-6}(\boldsymbol{x}_t)\}$,on the PES, e.g.,

$$\boldsymbol{a}_t = \pm(0, \dots, 0, c, 0, \dots, 0) = \pm c\boldsymbol{v}_i, \quad (17)$$

The step size, $c$, is fixed to 5.0 degree in our model. After taking an action $\boldsymbol{a}_t$ in the state $\boldsymbol{s}(\boldsymbol{x}_t)$, the agent moves to $\boldsymbol{x}_{t+1} = \boldsymbol{x}_t + \boldsymbol{a}_t$. Note that if the search space is restricted to a subspace, such as the dihedral angles in this study, constrained optimization must be performed to obtain the molecular geometry corresponding to $\boldsymbol{x}_{t+1}$, relaxing the remaining degrees of freedom while holding the updated RL coordinates fixed. Then, receives a new state $\boldsymbol{s}(\boldsymbol{x}_{t+1})$ defined by eqs. (14) and (15). Simultaneously, based on the new state, $\boldsymbol{s}(\boldsymbol{x}_{t+1})$, the agent also receives a reward,

$$r_t = \begin{cases} 10 & \text{if the agent reaches a TS region} \\ -0.01 & \text{otherwise} \end{cases}. \quad (18)$$

Here the TS regions are defined as the set of the points $\{\boldsymbol{x}\}$ satisfying eq. (8), where the root-mean-square value of gradient is below a threshold,

$$|\boldsymbol{g}(\boldsymbol{x})|/\sqrt{3N-6} \leq thrsh \quad (19)$$

The specific value of $thrsh$ is provided in section 3.1. Each episode terminates immediately if the agent reaches a TS region. In addition, an episode is truncated if the step count exceeds a predefined maximum, which is set to 1000 in all our DRL runs. This step limitation prevents the agent from exploring irrelevant regions and accumulating uninformative experiences.

**2-4. Bidirectional Long Short-Term Memory Network for variable-sized environments**

In this section, we describe the specific architecture used in this study for the Q function, $Q_\theta(\boldsymbol{s}, \boldsymbol{a})$. As mentioned in the Introduction, our previous study [37] developed a DQN model limited to fixed dimensional systems, where $Q_\theta(\boldsymbol{s}, \boldsymbol{a})$ was approximated using a single multilayer perceptron (MLP) as shown in Figure S1 (a). To apply our approach to molecular systems, it is essential to extend our model to handle variable-sized inputs. For this purpose, we developed a novel recurrent architecture combining an MLP and a bidirectional long short-term memory (BiLSTM) network, which was originally designed for processing sequential data in phoneme classification and natural language processing (NLP) tasks [50-52], as shown in Figure S1(b). The compact and fully unrolled representations of this novel architecture are illustrated on the left and right sides of Figure 1, respectively.

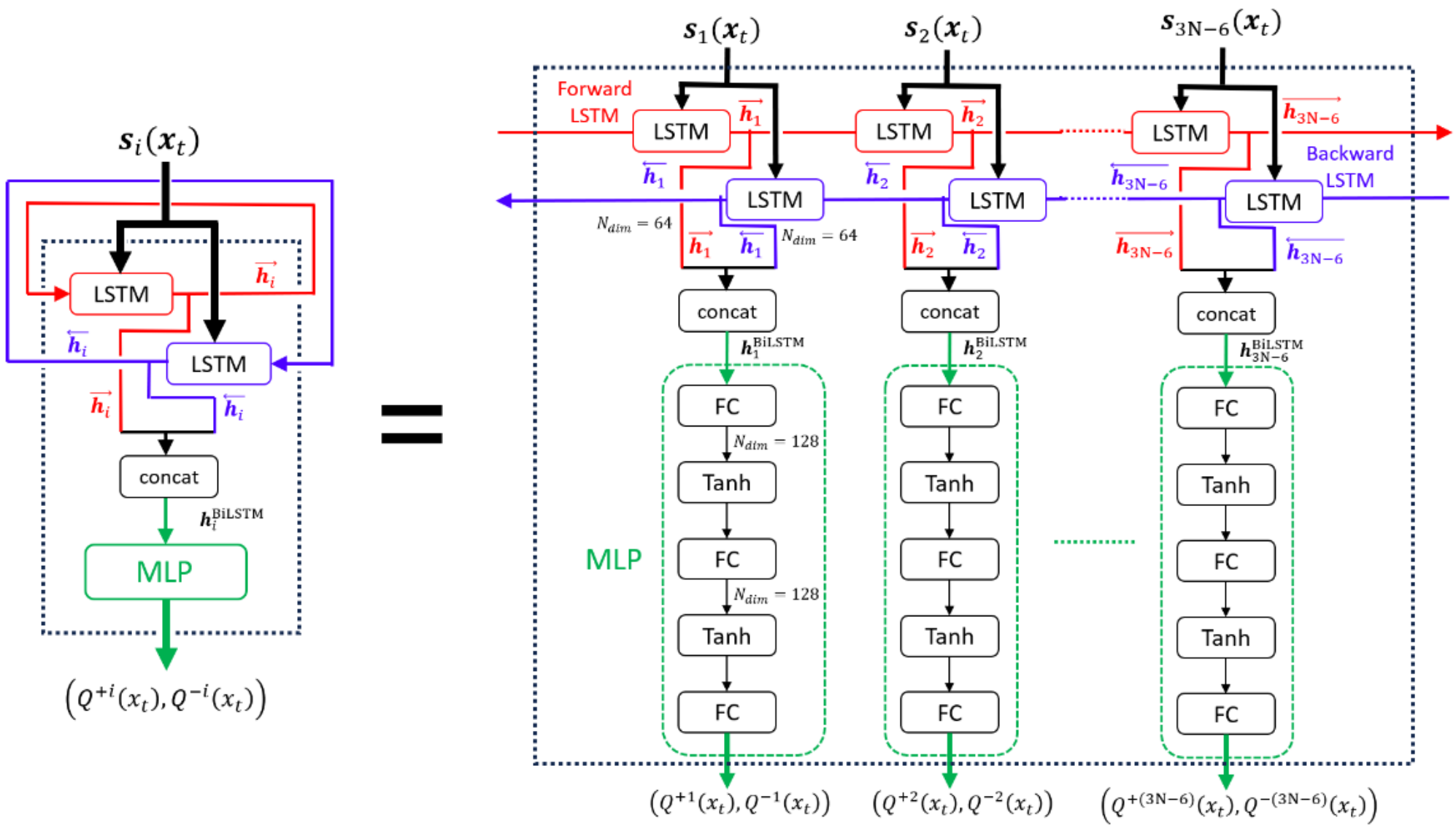

**Figure 1** Representation of the Q-function architecture employed. The compact form and unrolled form are presented on the left and right sides of the figure, respectively. The forward and backward long short-term memory (LSTM) networks and the multilayer perceptron (MLP) are colored in red, purple, and green, respectively. The outputs of the BiLSTM layer are concatenated into a single vector, $\boldsymbol{h}_i^{\mathrm{BiLSTM}}$, which is then fed into the MLP. The MLP block consists of three fully connected (FC) layers with hyperbolic tangent (Tanh) activation functions. The dimensionalities of the hidden states and the FC outputs are denoted by $N_{\mathrm{dim}}$.

The BiLSTM consists of forward and backward LSTMs (colored in red and purple in Figure 1, respectively) that process sequences in both directions to effectively extract contextual dependencies among the information units, analogous to syntactic relationships in NLP [53-55]. In our model, the information unit corresponds to the feature vector of a vibrational mode, $s_i(\boldsymbol{x}_t)$; sequentially feeding these vectors enables the network to bidirectionally transfer information across low- and high-frequency modes, yielding forward and backward hidden states, $\overrightarrow{\boldsymbol{h}_\iota}$ and $\overleftarrow{\boldsymbol{h}_\iota}$

$$\overrightarrow{\boldsymbol{h}_\iota} = \mathrm{LSTM}(s_i(\boldsymbol{x}_t), \overrightarrow{\boldsymbol{h}_{\iota-1}}), \quad (20)$$

$$\overleftarrow{\boldsymbol{h}_\iota} = \mathrm{LSTM}(s_i(\boldsymbol{x}_t), \overleftarrow{\boldsymbol{h}_{\iota+1}}). \quad (21)$$

The output of the BiLSTM layer is then obtained by concatenating $\overrightarrow{\boldsymbol{h}_\iota}$ and $\overleftarrow{\boldsymbol{h}_\iota}$:

$$\mathrm{BiLSTM}\big(s_i(\boldsymbol{x}_t)\big) = \boldsymbol{h}_i^{\mathrm{BiLSTM}} = \overrightarrow{\boldsymbol{h}_\iota} \oplus \overleftarrow{\boldsymbol{h}_\iota} \quad (22).$$

This vector encodes the comprehensive dependencies among all vibrational modes in both directions. Subsequently, a multilayer perceptron (MLP) is adopted to estimate Q-values for the positive and negative directions of the i-th vibrational mode based on $\boldsymbol{h}_i^{\mathrm{BiLSTM}}$:

$$\left(Q_\theta^{\pm i}(\boldsymbol{x}_t)\right) \equiv \left(Q_\theta^{+i}(\boldsymbol{x}_t), Q_\theta^{-i}(\boldsymbol{x}_t)\right) = \mathrm{MLP}\left(\boldsymbol{h}_i^{\mathrm{BiLSTM}}\right). \quad (23)$$

In summary, the Q-function is formulated as,

$$Q_\theta(\boldsymbol{s}, \boldsymbol{a}) = \left(Q_\theta^{\pm 1}(\boldsymbol{s}(\boldsymbol{x}_t)), Q_\theta^{\pm 2}(\boldsymbol{s}(\boldsymbol{x}_t)), \dots, Q_\theta^{\pm(3\mathrm{N}-6)}(\boldsymbol{s}(\boldsymbol{x}_t))\right), \tag{24}$$

where the Q-value of each action, $a^j$, is given by,

$$Q_\theta(\boldsymbol{s}, a^j) = Q_\theta^j(\boldsymbol{s}(\boldsymbol{x}_t)) = \mathrm{MLP}\left(\mathrm{BiLSTM}\left(s_{|j|}(\boldsymbol{x}_t)\right)\right), j = \pm 1, .., \pm(3\mathrm{N}-6). \tag{25}$$

The entire training procedure of our DRL model is summarized in Algorithm 1.

**Algorithm 1 Training procedure of the DQN model for TS searches**

1: Initialize the action-value function $\boldsymbol{Q_\theta}$ (both Multilayer perceptron and BiLSTM)
2: Initialize replay buffer $\boldsymbol{\mathcal{D}}$ and total time step $\boldsymbol{t = 0}$
3: for episode = 1 to T do
4: Initialize the termination flag, TERM=.False.
5: Initialize agent's position $\boldsymbol{x_0}$
6: Calculate Hessian and gradient to get the initial state $\boldsymbol{s_0}$
7: while TERM != .True. do
8: Feed the state to the Q-network $\boldsymbol{Q_\theta(s, a)}$
9: Select an action $\boldsymbol{a_t}$ following the policy $\boldsymbol{\pi_\varepsilon(s, a)}$
10: Update the internal coordinate $\boldsymbol{x_{t+1}}$ in the direction corresponding to the action $\boldsymbol{a_t}$
11: Perform constrained optimization with the updated $\boldsymbol{x_{t+1}}$.
12: Calculate the Hessian and gradient to get the next state $\boldsymbol{s_{t+1}}$
13: Get a reward $\boldsymbol{r_t}$
14: If any termination condition is satisfied, then set TERM=.True.
15: Store the experience $\{\boldsymbol{s_t}, \boldsymbol{a}(\boldsymbol{x_t}), \boldsymbol{r_t}, \boldsymbol{s_{t+1}}\}$ into replay buffer $\boldsymbol{D}$
16: If (t mod 10) == 0 then sample a mini-batch from $\boldsymbol{D}$ and train Q-network
15. If (t mod 1000) == 0 then copy Q-network $\boldsymbol{Q_\theta}$ to target Q-network, $\boldsymbol{Q_\theta^{target}}$
16. $\boldsymbol{t = t + 1}$
17. end while
18.end for
19.return

## 3. Results and Discussion

### 3.1 Application to organic molecular systems: molecular modeling and computational setup

We selected 1,3,5-hexatriene (Triene), alanine dipeptide (Ala2), and 1,3,5,7-octatetraene (Tetraene) [Figures 2(a)–(c)] as model systems, all of which consist of structurally rigid planar units linked by rotatable bonds. Because the conformational flexibility of these systems is governed by torsional rotations, and given that our DRL approach involves computationally intensive trial-and-error exploration, we restricted the search space in each DRL step to the backbone dihedral angles: $(\phi, \psi)$ for Triene and Ala2, and $(\phi, \psi, \theta)$ for Tetraene. In figure 2 (a)-(c), we illustrated their most stable, all-*trans* (or fully extended) conformations, in which all defined dihedral angles are 180° in accordance with IUPAC conventions [56,57]. Since the goal of the calculations described below is to evaluate not only whether our model can learn the TS search task in molecular systems, but also whether it possesses robust transfer learning capabilities, the order in which these systems are trained is of crucial importance. Accordingly, we first trained our model on 1,3,5-hexatriene (Triene)—the simplest system among them—as the source system from which knowledge is transferred [58–65], and subsequently applied the resulting model to Ala2 and Tetraene to examine transferability across target tasks involving backbone modification and size expansion.

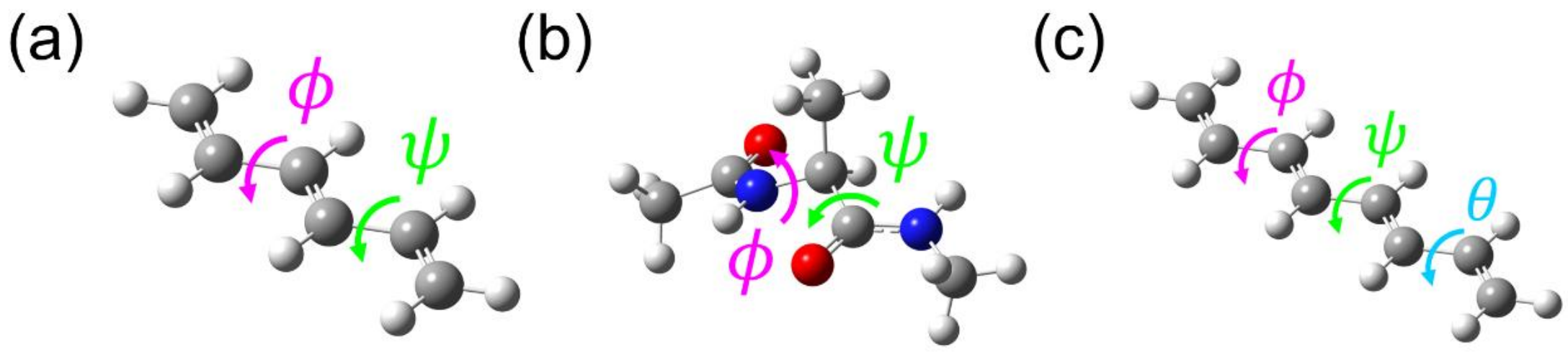

**Figure 2.** Molecular structures and defined dihedral angles: (a) 1,3,5-hexatriene (Triene; $\phi, \psi$), (b) alanine dipeptide (Ala2; $\phi, \psi$), (c) 1,3,5,7-octatetraene (Tetraene; $\phi, \psi, \theta$). All structures are shown in their most stable, all-*trans* (or fully extended) conformations.

To facilitate knowledge transfer across the different molecular systems, the computational settings were largely common to all systems. The step size of the DRL runs for all molecules is 5.0 degrees. The RMS gradient threshold in eq. (19) for identifying the TS regions is set to 0.001 Hartree/rad for all molecules. This convergence criterion is looser than the standard criteria for TS optimization (e.g., $3.0 \times 10^{-4}$ Hartree/rad in the default settings of Gaussian 09), resulting in broader target TS regions, but this setting is intentional: due to the inherent stochasticity of RL exploration, pinpointing precise TS geometries solely through RL runs is impractical. To compensate for this loose threshold, we confirmed that subsequent standard local optimization procedures, such as the Newton-Raphson method, successfully converge to the precise TS points starting from the geometries that satisfy this condition for identifying the TS regions. All calculations of gradients and curvatures necessary for the DRL steps and subsequent TS optimizations were carried out using the Gaussian 09 (Rev. C.01) software package [66]. In these calculations, only the force field varied depending on the system: the Universal Force Field (UFF) [67] was employed for Triene, and Tetraene, whereas the Amber ff94 force field was used for Ala2 [68]. The other hyperparameters for the DRL runs used in this study are also common across the systems, as summarized in Table S1.

### 3.2. 1,3,5-hexatriene (Triene): Validation of the DRL model

We first present the results for 1,3,5-hexatriene (Triene), which serves as the foundational system for the learning stages in this study. Figure 3 (a) shows the PES contour map of Triene, representing the minimized energy projected onto the DRL search space of the two backbone dihedral angles, $(\phi, \psi)$ [Figure 2(a)], where the EQ and TS regions are colored in red and green, respectively. For consistency, this projected PES representation and the same color scheme for the EQ and TS regions are adopted in all subsequent subsections for the other systems. Note that this contour map displays only the stationary points that satisfy Eq. (8) upon dimensional reduction, excluding several stationary regions of the full system that fail to meet this condition.

The exact coordinates of the EQ and TS points for Triene, along with their relative enegies, are provided in Table S2. As shown in Figure 3 (a), the PES exhibits a simple grid-like landscape, reflecting that both $\phi$ and $\psi$ possess potential energy minima at 0° and 180°, corresponding to cis (C) and trans (T) conformations, and a maximum at $\pm 90°$,in agreement with a previous *ab initio* study of 1,3,5-hexatriene [69]. Consequently, the EQ regions are identified by combinations of these minima, i.e., $(\phi, \psi) \sim (0°, 0°), (0°, 180°), (180°, 0°),$ and $(180°, 180°)$, corresponding to cis-cis (CC), cis-trans (CT), trans-cis (TC), and trans-trans (TT), respectively. In contrast, the TS regions are characterized by a combination of a minimum and a maximum, i.e., $(\phi, \psi) \sim (0°, \pm 90°)$, $(180°, \pm 90°), (\pm 90°, 0°),$ and $(\pm 90°, 180°)$. A minor complication is that two distinct EQs exist in each of the regions around $(\phi, \psi) \approx (0°, 0°), (0°, 180°)$, and $(180°, 0°)$. This arises from steric repulsion between hydrogen atoms on adjacent planar units in the cis conformation (i.e. where either $\phi$ or $\psi$ is $0°$, or both)

Due to the simple landscape of the PES, the reaction pathways can also be readily mapped out via the steepest descent paths from the TS regions to the EQs, as illustrated in Figure 3 (b). Note that because the starting points are TS regions satisfying Equations (8) and (19) rather than exact TS points, the terms "reaction pathway" and "minimum energy path" (MEP) used in our RL analysis differ from conventional definitions, where each TS point connects exactly two EQs.

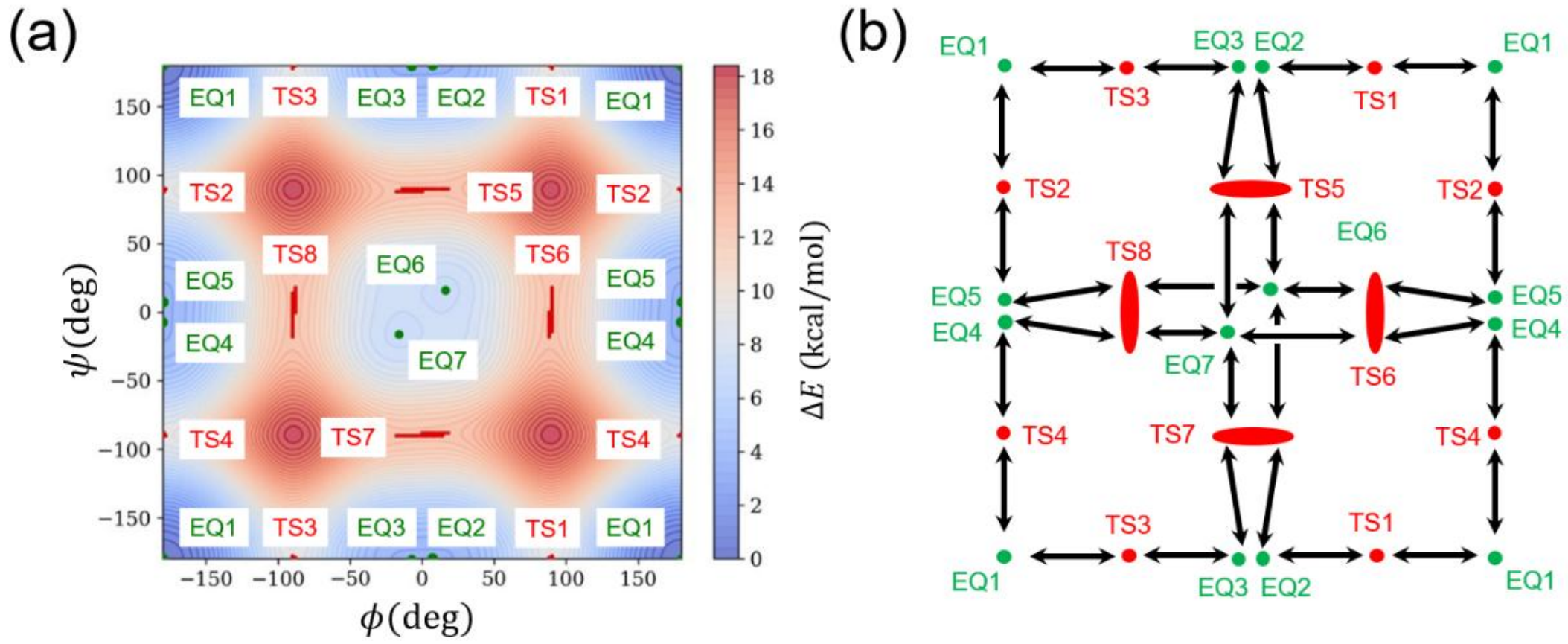

**Figure 3.** (a) Potential energy surface (PES) contour map of 1,3,5-hexatriene as a function of the backbone dihedral angles $(\phi, \psi)$. Equilibrium (EQ) and transition state (TS) regions are highlighted in green and red, respectively. The coordinates of these EQ and TS points are summarized in Table S2. (b) Schematic representation of the reaction pathways (MEPs) connecting the TS regions to the EQs, indicated by double-headed black arrows.

We examined the performance of the RL agent on this system under the computational setup summarized in Table S1 (common to all systems). Specifically, training was carried out for up to 1000 rounds, each comprising one episode originating from each EQ point. We evaluated the mean return for each round, defined as $\Sigma_{i=1}^{N_{EQ}} \Sigma_t r_t / N_{EQ}$, where $N_{EQ}$ is the number of EQ points on the PES. Furthermore, to reduce the dependence on the initial parameters of the Q-network and the stochastic exploration process, the learning curves presented here are averaged over 10 independent runs. The same evaluation procedures were also applied to the other molecular systems described in the following subsections. The resulting learning curve and its 10-round moving average are shown in Figure 4 (a) and (b), respectively. Although some fluctuations arising from stochastic actions for exploration occurred in the early stages, our model successfully learned the TS search task for Triene, consistent with observations in the model systems of our previous study [43]. Specifically, the average return increased rapidly and nearly saturated at approximately 400 rounds. To quantitatively evaluate the learning time, we introduced two thresholds of 8.0 and 9.0, corresponding to two proficiency levels, “intermediate” and “advanced”, respectively. Based on the 10-round moving average of the learning curve shown in Figure 4 (b), the agent reached the intermediate level at round 91 (episode 637), indicating that it, on average, reached a TS regions within 200 steps. With further training, the agent progressed to the advanced level at round 284 (episode 1988), indicating that it located one of the TS regions within 100 steps and successfully acquired the effective strategy for finding TS structures.

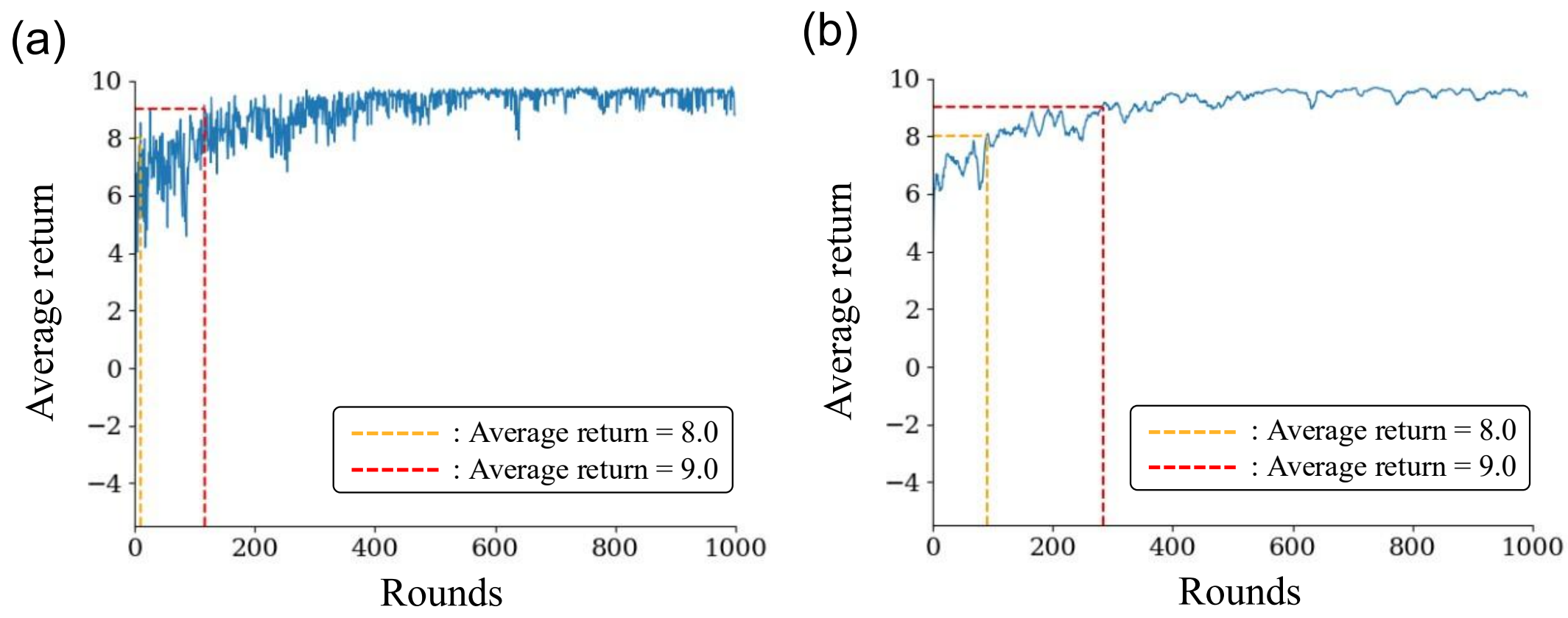

**Figure 4.** Learning curves of the DRL model for Triene: (a) average return over 10 independent runs with random initializations and (b) 10-round moving averages of the curves in (a). Orange and red dashed lines indicate the return thresholds of 8.0 (intermediate level) and 9.0 (advanced level), respectively.

**Table 1.** Percentages of trajectories reaching each TS region for Triene, categorized by the starting EQ (the corresponding counts are listed in Table S3). Asterisks indicate the TS regions connected to each EQ via MEPs, and their total percentages are provided in the bottom row.

| | EQ1(TT) | EQ2(CT) | EQ3(CT) | EQ4(TC) | EQ5(TC) | EQ6(CC) | EQ7(CC) |
|---|---|---|---|---|---|---|---|
| TS1 | 6.56* | 1.19* | 0.10 | 0.06 | 0.04 | 0.02 | 0.06 |
| TS2 | 25.49* | 0.12 | 0.04 | 0.06 | 0.34* | 0.00 | 0.06 |
| TS3 | 26.74* | 0.10 | 0.26* | 0.24 | 0.02 | 0.02 | 0.00 |
| TS4 | 6.28* | 0.04 | 0.04 | 0.85* | 0.18 | 0.00 | 0.00 |
| TS5 | 13.15 | 77.89* | 79.52* | 0.26 | 0.04 | 20.08* | 0.32* |
| TS6 | 4.37 | 0.16 | 0.16 | 20.51* | 20.04* | 78.61* | 1.14* |
| TS7 | 3.83 | 20.35* | 19.84* | 0.20 | 0.18 | 0.96* | 78.33* |
| TS8 | 13.57 | 0.14 | 0.04 | 77.81* | 79.16* | 0.30* | 20.09* |
| Total for TS regions connected via MEPs | 65.08 | 99.54 | 99.72 | 99.23 | 99.72 | 99.96 | 99.88 |

To examine the TS search paths traced by the agent, Table 1 summarizes the percentages of trajectories that reached each TS region starting from each EQ point (the corresponding trajectory counts are listed in Table S3). Asterisks denote the TS regions connected to each EQ via the steepest descent paths (see Figure 3(b)) and the total percentages of trajectories reaching these TS regions are summarized at the bottom line of the table. These totals reveal that the agent predominantly reached the TS regions connected via the minimum energy paths (MEPs), which are obtained as the steepest descent paths from TS regions. This implies that the agent autonomously learned a TS search strategy intrinsically based on local gradients, curvatures, and changes in curvature, which quantum chemists have developed over more than half a century. Particularly for EQ2–EQ7, almost all trajectories (>99%) reached the TS regions connected via the steepest descent paths (i.e., the MEPs). The sole exception was EQ1, from which 65.1% of the trajectories reached the connected TS regions (TS1–TS4), with the remainder deviating to unconnected TS regions. This is likely due to the fact that the regions for TS1–TS4 are considerably smaller than those for TS5–TS8 as shown in Figure 2(a); consequently, due to the stochastic nature of its exploration t on the PES, the agent tended to miss the nearby yet small TS regions and instead encountered the larger TS regions (TS5–TS8).

Figures 5 (a)-(c) shows the representative trajectories at the first round, intermediate level, and advanced level, respectively. These plots indicate a clear improvement as learning progressed. This change in trajectory is reasonable: in the early stages, the agent explored various directions randomly and thus often entered unproductive regions, including high-energy regions, as shown in Figure 5(a). Indeed, in the initial round, the agent reached a TS region in only

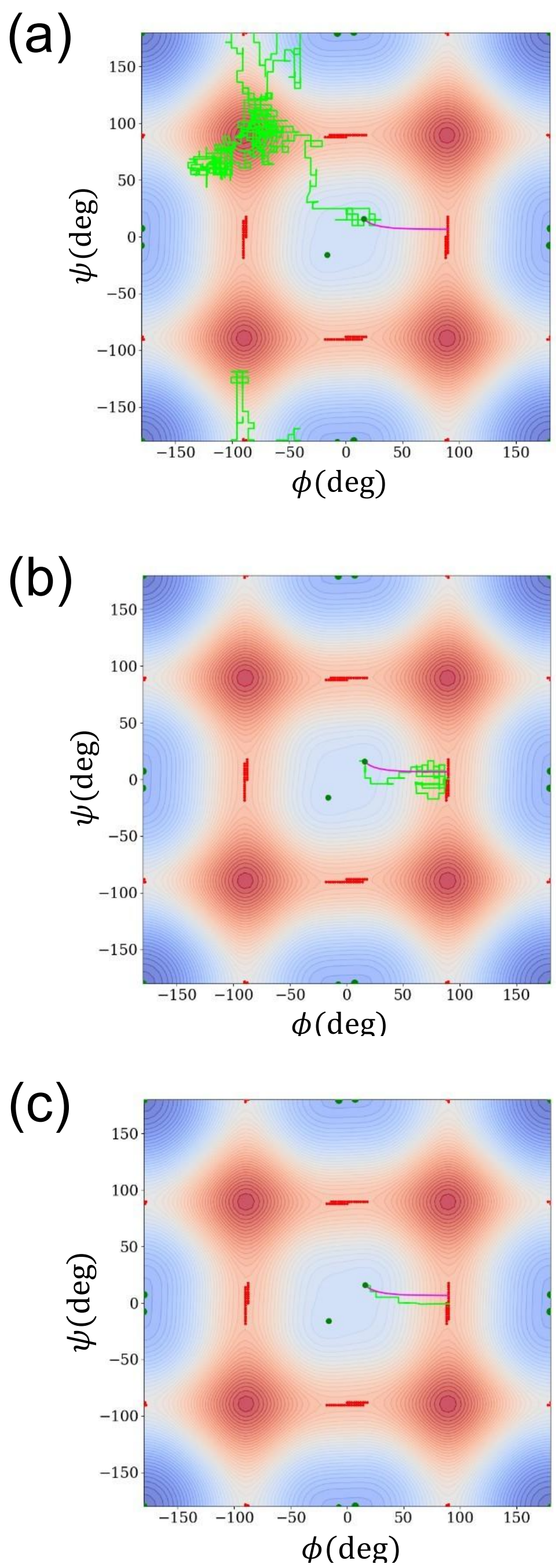

**Figure 5.** Representative trajectories on the PES of Triene at the (a) first round, (b) intermediate, and (c) advanced levels. Red points, green points, and lime-green lines denote the EQs, TSs, and agent trajectories, respectively. The steepest descent paths originating from TS6 are shown in magenta.

11 of the 35 trials (five trials from each of the seven EQs). In contrast, Figure 5(b) demonstrates that the agent at the intermediate level avoided such unfavorable regions, and successfully reached the TS regions. As the learning progressed, the agent learned to navigate efficient pathways to the nearby TS regions via the MEPs, as illustrated in Figure 5 (c). In fact, as shown in Figure 5 (c), the trajectory of the well-trained agent closely follows the steepest descent path (i.e., the MEP) depicted in magenta, which is fully consistent with the results in Table 1. It is noteworthy that nearly all trajectories of the advanced agent whose average return exceeds 9.0 similarly trace the MEPs, reaching the TS regions within 100 steps.

The computational results described in this section indicate that our model successfully learned a strategy for TS search in Triene using local PES features. In the context of transfer learning [58-65], the TS search on a simple PES landscape (Figure 3 (a)) and reaction pathways (Figure 3 (b)), as presented in this subsection, represents a "source task" that serves as the foundation for tackling more complex problems, i.e., "target tasks". In the following sections, we apply the model trained in this section to alanine dipeptide (Section 3.4) and to tetraene (Section 3.5) to examine the transfer-learning effects of our model.

### 3.3. Alanine dipeptide: Transferability to backbone modification

In this subsection, we applied our model trained on Triene to alanine dipeptide (Ala2) to test the transferability upon backbone modification. Alanine dipeptide (Ala2) consists of two peptide planes rigidified by partial conjugation, sharing with Triene the presence of multiple stationary points across the two-dimensional dihedral angle space $(\phi, \psi)$. Nevertheless, notable structural distinctions exist between the two systems, including the elemental composition of the molecular backbone and the hinge connecting the rigid units—namely, an $sp^3$-hybridized $\alpha$-carbon in Ala2 versus carbon–carbon single bonds in Triene. In fact, the PES landscape of Ala2 as a function of the dihedral angles $(\phi, \psi)$ [Figure 6(a)] is considerably different from that of Triene shown in Figure 3(a). Two additional spurious TS regions arising as artifacts of the coarse conditions are omitted here (see Figure S2 and its caption). The exact locations and energies of the EQ and TS points, along with their corresponding geometries, are provided in Table S4. Ala2 possesses fewer EQs and TSs than Triene, and its reaction pathway network is considerably distorted as a consequence of the highly skewed PES landscape, as shown in Figures 6 (a) and 6 (b). Thus, the Ala2 system serves as an ideal testbed to assess the transferability of our Triene-trained model to backbone modifications.

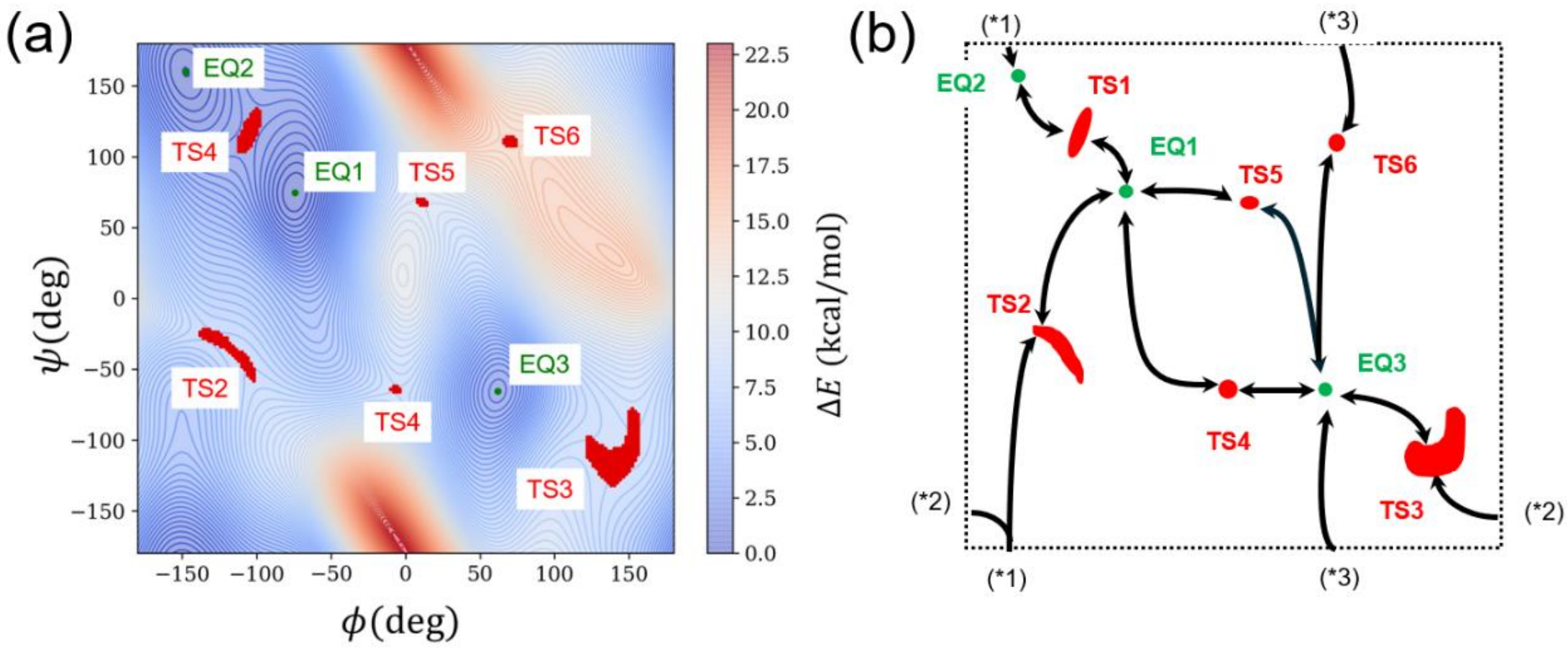

**Figure 6.** (a) Potential energy surface (PES) contour map of alanine dipeptide (Ala2) as a function of the backbone dihedral angles $(\phi, \psi)$. Equilibrium (EQ) and transition state (TS) regions are highlighted in green and red, respectively. The coordinates of these EQ and TS points are summarized in Table S4. (b) Schematic representation of the reaction pathways (MEPs) connecting the TS regions to the EQs, indicated by double-headed black arrows. Matching asterisked labels (*1-*3) in (b) denote the pathways that continue across the periodic boundaries at $\phi = \pm 180^{\circ}$ and $\psi = \pm 180^{\circ}$.

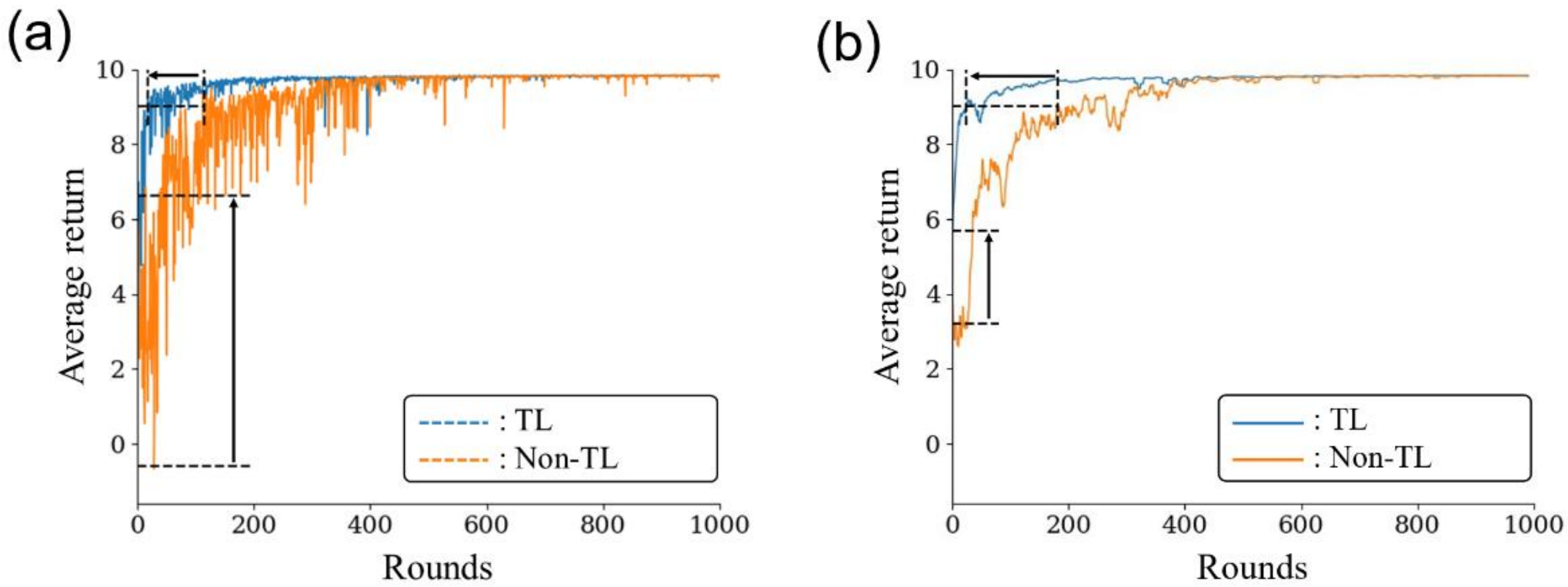

**Figure 7.** Learning curves of the DRL model with transfer-learning (TL) and without it (Non-TL) for alanine dipeptide (Ala2): (a) average return over 10 independent runs with random initializations and (b) 10-round moving averages of the curves in (a). The dashed lines mark the initial return values and the threshold for the advanced level (9.0). The vertical black arrows indicate the jump-start effect, and the horizontal black arrows represent the reduction in learning time required to reach the advanced level.

Then, we conducted DRL runs on Ala2 using two distinct agents: one pre-trained on the Triene TS search task for 1,000 rounds in the previous section (denoted as "TL") and another trained from scratch with no prior learning (denoted as "Non-TL"). Figures 7(a) and 7(b) present the learning curves and their 10-round moving averages, respectively. These plots clearly demonstrate two distinct transfer effects: the jump-start and learning-time reduction effects [58-62], consistent with our previous study on a two-dimensional model system [43]. Based on the 10-round moving averages shown in Figure 7(b), we evaluated the initial average return and the number of rounds (with episodes in parentheses) required to reach the intermediate and advanced performance levels, which are summarized in Table 2. Specifically, the initial average return increased from 3.20 to 5.70, suggesting that the TL agent had already adopted an effective search strategy at a very early stage, acquired through its prior experience on Triene. Furthermore, the numbers of rounds required to reach the intermediate and advanced levels decreases drastically from 109 to 9, and 182 to 25, respectively. For comparison with the later applications to other systems with different number of EQs, the corresponding numbers of episodes are also shown in parenthesis. These data demonstrate that once trained on a simple system, our model can rapidly learn the TS search task in a distinct system, even when the PES landscape of the target task (Ala2 in this case) is substantially distorted relative to that of the source task (Triene) due to the backbone modifications.

**Table 2.** The initial average returns and the number of rounds required to reach the intermediate and advanced levels for the Non-TL and TL agents. Values in parentheses for the latter two metrics denote the corresponding numbers of episodes.

| Agent | Initial average return | Intermediate level [round (episode)] | Advanced level [round (episode)] |
|---|---|---|---|
| Non-TL | 3.20 | 109 (327) | 182 (546) |
| TL | 5.70 | 9 (27) | 25 (75) |

Table 3 summarizes the percentages of trajectories that reached each TS region starting from each EQ point for the Non-TL and TL agents (the corresponding trajectory counts are listed in Table S5). Asterisks denote the TS regions connected to each EQ via MEPs (see Figure 6(b)), and the total percentages of trajectories reaching these TS regions are summarized in the bottom row of the table. For both the Non-TL and TL agents, over 90% of the trajectories successfully reached the TS regions directly linked to the respective starting EQs (Figure 3(b)), with the majority concentrating on the nearest TS regions. Notably, for the TL agent, nearly 99% or more of the trajectories terminated in the MEP-connected TS regions.

**Table 3.** Percentages of trajectories reaching each TS region for Ala2, categorized by the starting EQ (the corresponding counts are listed in Table S5). Asterisks indicate the TS regions connected to each EQ via MEPs, and their total percentages are provided in the bottom row.

| | EQ1 | | EQ2 | | EQ3 | |
|---|---|---|---|---|---|---|
| | Non-TL | TL | Non-TL | Non-TL | TL | Non-TL |
| TS1 | 91.34 * | 85.29 * | 88.41 * | 96.63 * | 0.33 | 0.04 |
| TS2 | 3.58 * | 1.30 * | 1.77 | 0.69 | 1.09 | 0.10 |
| TS3 | 0.24 | 0.04 | 3.45 * | 2.32 * | 76.87 * | 88.32 * |
| TS4 | 1.09 * | 2.24 * | 0.16 | 0.06 | 11.10 * | 8.65 * |
| TS5 | 2.35 * | 10.84 * | 0.12 | 0.02 | 3.19 * | 1.65 * |
| TS6 | 1.40 | 0.28 | 6.08 | 0.28 | 7.42 * | 1.24 * |
| Total for TS regions connected via MEPs | 98.36 | 99.68 | 91.86 | 98.85 | 99.58 | 99.86 |

To confirm the improvement in the agent's proficiency, Figures 8 (a)–(f) show representative trajectories starting from EQ2, along with the steepest descent path (i.e., the MEP) originating from TS3, at the first-round, intermediate, and advanced levels for the Non-TL (Figures 8 (a)–(c)) and TL (Figures 8 (d)–(f)) agents, respectively. The trajectory of the Non-TL agent in the first round barely reached TS1, as shown in Figure 8 (a), reflecting that the TS search success rate of the Non-TL agent was 60% (9 successes and 6 failures across 15 trials from the three starting EQs) with an average of 680 steps required to reach a TS region. As training progressed, similar to the trajectories for Triene shown in Figure 5(b),(c), the trajectories of the Non-TL agent steadily converged toward the MEP as shown in Figure 8 (b) and (c). In contrast, even in the first round (Figure 8(d)), the trajectories of the TL agent exhibited an efficiency comparable to—albeit slightly lower than—that of the Non-TL agent at the intermediate level, as reflected by an initial average return exceeding 6.0 (Figure 7(a)). Subsequently, the TL trajectories rapidly converged toward the MEP, reaching the intermediate and advanced levels at rounds 9 and 25, respectively (Figures 8(e) and 8(f)). This indicates that the jump-start and learning-time reduction effects of transfer learning shown in Table 2 are also manifested in the trajectory behaviors.

Overall, the results in this section demonstrate that our model pre-trained on Triene (the TL agent) achieved distinct transfer learning benefits compared to the Non-TL agent. Given that the PES landscapes of Triene and Ala2 differ substantially, as shown in Figures 3 (a) and 6 (a), this suggests that the agent has acquired a system-independent exploration strategy. Combined with our previous findings on model systems, our model is expected to accelerate learning in different molecular systems, provided that they share the same dimensionality. In the following section, we examine the remaining question: whether the agent exhibits transferability across different dimensions as well.

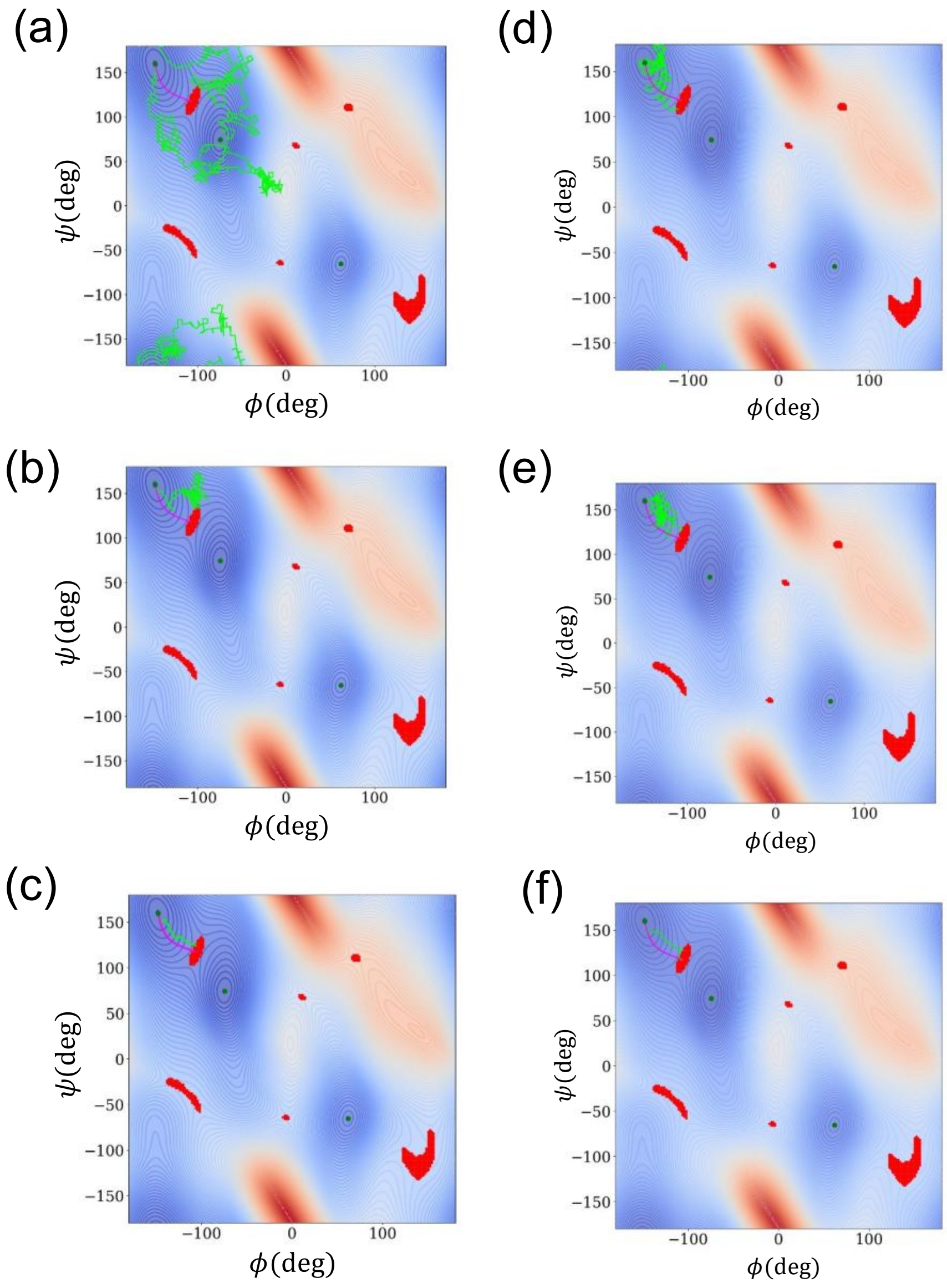

**Figure 8.** Representative trajectories along with the number of steps for $Ala_2$: (a) at the initial round, (b) at the intermediate level, and (c) at the advanced level for the Non-TL agent; (d) at the initial round, (e) at the intermediate level, and (f) at the advanced level for the TL agent. Green points, red points, lime-green lines, and magenta lines denote the starting EQ points, TS regions, and agent trajectories, the MEPs originating from TS3.

### 3.4. 1,3,5,7-octatetraene: Transferability to Chain Extension

To evaluate whether the transferability of our model extends to higher-dimensional systems, we applied the agent trained on Triene to a longer polyene, specifically 1,3,5,7-octatetraene (Tetraene) in this subsection. Tetraene has 17 equilibrium structures (EQs) and 38 transition structures (TSs) along the backbone dihedrals, whose coordinates and relative energies with respect to the most stable EQ are summarized in Table S6.

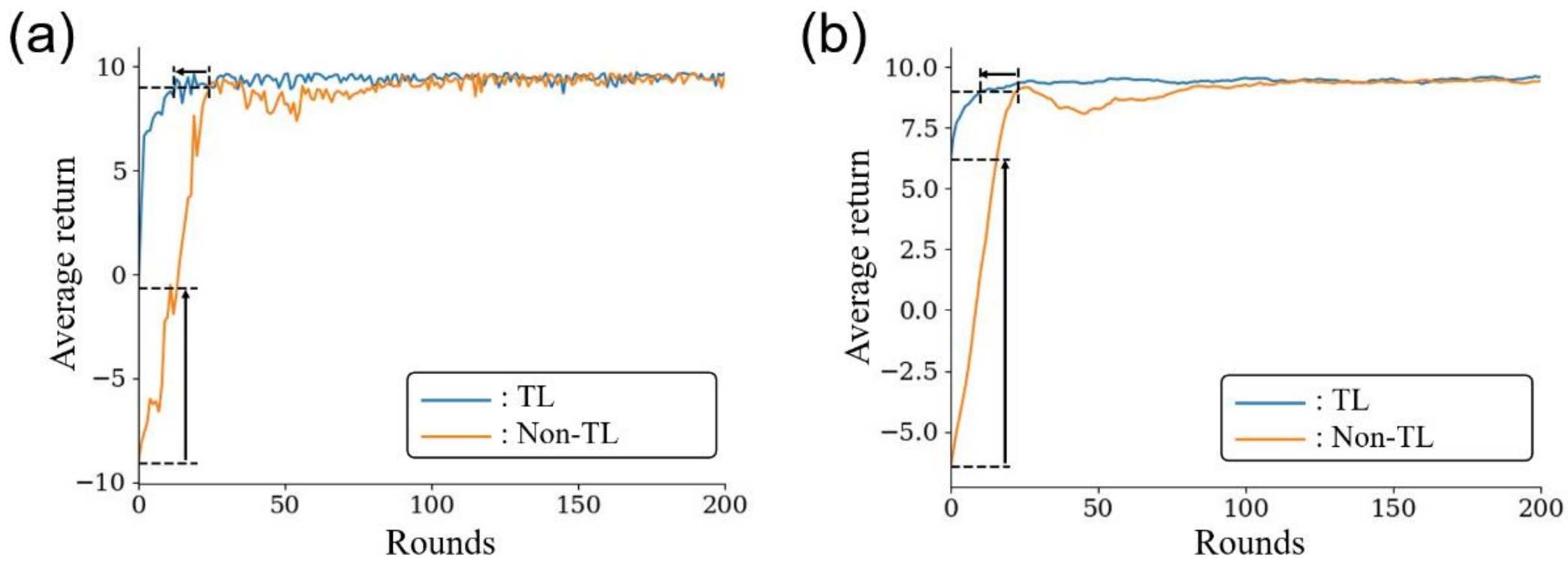

**Figure 9.** Learning curves of the DRL model over the first 200 rounds with transfer-learning (TL) and without it (Non-TL) for 1,3,5,7-octatetraene (Tetraene): (a) average return over 10 independent runs with random initializations and (b) 10-round moving averages of the curves in (a). The dashed lines mark the initial return values and the threshold for the advanced level (9.0). The vertical black arrows indicate the jump-start effect, and the horizontal black arrows represent the reduction in learning time required to reach the advanced level.

**Table 4.** The initial average returns and the number of rounds required to reach the intermediate and advanced levels for the Non-TL and TL agents. Values in parentheses for the latter two metrics denote the corresponding numbers of episodes.

| Agent | Initial average return | Intermediate level [round (episode)] | Advanced level [round (episode)] |
|---|---|---|---|
| Non-TL | -6.46 | 20 (340) | 24 (408) |
| TL | 6.22 | 5 (85) | 12 (204) |

We conducted DRL runs on tetraene using both the TL agent (pre-trained on the Triene TS search task for 1,000 rounds) and the Non-TL agent (without prior training), as in the case of Ala2. Figures 9 (a) and (b) shows the learning curves and their 10-round moving averages, respectively. For clarity, only curves for the first 200 rounds are shown (see Figure S3 for the full curves over 1,000 rounds).  Once again, these plots clearly demonstrate both the jump-start effect and the reduction in training time.  Table 4 lists the initial average returns and the number of rounds (episodes in parentheses) needed to reach the intermediate and advanced performance levels, both derived from the 10-round moving averages in Figure 9 (b).  These data show that the initial mean return improves from −6.46 to 6.22, representing a drastic increase of over 12, and that the number of rounds (episodes) required to reach the intermediate and advanced levels is also reduced by a factor of 4 and 2, respectively.  In summary, these findings demonstrate that our DRL model, incorporating a BiLSTM+MLP architecture for variable-sized systems, operates successfully under dimensional expansion, thereby transferring learned local policy features to systems with additional degrees of freedom.

Considering that our transfer scheme relies solely on the BiLSTM to accommodate variable-sized molecular systems without any reward shaping or task mapping [58–65], such success is particularly remarkable.  In a theoretical framework for transfer learning in a zero-shot setting formalized by Kirk et al. [64], transferring the TS search task from Triene to Tetraene corresponds to extrapolation along an ordinal axis, which reflects the capacity to extrapolate learned policies to systems with higher dimensionalities and extended degrees of freedom.  As discussed in their study, because no general-purpose algorithm can guarantee arbitrary extrapolation across tasks, tackling such extrapolative regimes requires stronger inductive biases targeted toward specific types of variation [64].  For instance, using a graph attention network (GAT)+MLP framework, Garg et al. demonstrated that a model pre-trained on small graphs substantially outperformed models trained from scratch across benchmark domains (SysAdmin, Game of Life, and Academic Advising) without architectural modifications or fine-tuning, by leveraging pairwise local dependencies—such as network connectivity, adjacent grid cells, and course prerequisites—as a strong inductive bias to resolve global tasks through local contextual actions [65].  Remarkably, our BiLSTM+MLP architecture detailed above succeeded in achieving effective policy transfer across molecular systems with different dimensionalities, consistent with this underlying mechanism.  In our study, the shared local structural motifs of dihedral angle rotations of fixed conjugated planes, together with our model design grounded in established quantum-chemical (QC) TS search methods [11-25] incorporating potential energy surface topography, work as physical inductive biases that seamlessly bridge both backbone substitutions and dimensional scaling.  In fact, as shown in Table S7, almost all trajectories of the transferred agent navigated along the minimum energy paths (MEPs) from equilibrium points to the transition state regions.  This confirms that it successfully followed the direction along which the lowest curvature inverts from positive to negative, which is precisely the desired behavior of the deep reinforcement learning (DRL) approach combined with conventional quantum-chemical methods that we aimed to establish [11–25].

## Conclusions

In this study, we developed a novel deep reinforcement learning (DRL) approach to transition state (TS) search in molecular systems, combining the DQN algorithm with a BiLSTM network to satisfy two key conditions: (i) operating without prior information regarding either global potential energy surface landscapes or product geometries, and (ii) handling not only different molecules within the same dimensional search space but also systems of variable sizes. The framework is an extension of our previously developed DRL approach based on the DQN algorithm, in which we exploited gradients, curvatures, and changes of curvatures on 2D potential energy surfaces, inspired from established quantum-chemical TS search methods. Our method was successfully applied to learn TS search tasks for molecular systems, namely 1,3,5-hexatriene (Triene), alanine dipeptide (Ala2), and 1,3,5,7-octatetraene (Tetraene), by learning pathways from equilibrium points to TS regions along the minimum energy paths, analogous to the 2D PES models in our previous study [43]. More importantly, the extended method incorporating the BiLSTM architecture presented in this study exhibits two transfer-learning effects, i.e., jump-start performance and reduced learning time, both from Triene to Ala2 (backbone modification) and from Triene to Tetraene (chain extension). From the viewpoint of the transfer-learning theory, it is particularly noteworthy that our method transferred knowledge from the TS search task under chain extension across different dimensionalities without relying on reward shaping or task mapping, which fundamentally requires strong inductive biases [64,65]. In our DRL model, the established quantum-chemical TS search principle grounded in potential energy surface topography to navigate along the MEP toward TS regions works [11-25] as strong physical inductive biases.

Toward future applications to molecular systems, transferability across different molecular sizes and geometries represents a major strength of our approach, as it allows an agent trained on simpler systems to accelerate TS searches in larger or more complex ones. Thus, as the agent continues to learn, it can become increasingly versatile and proficient across diverse molecular systems. Although practical applications of deep reinforcement learning still face numerous challenges such as extensive trial-and-error runs, the stochastic nature of the exploration, and the curse of dimensionality associated with the search space, our findings on cross-system transferability in this study are expected to represent an important step forward in developing DRL-based approaches for TS searches.

## Acknowledgements

This work was supported by the Support for Pioneering Research Initiated by the Next Generation (SPRING) program of the Japan Science and Technology Agency (JST), Japan, (Grant Number JPMJSP2138); the Japan Society for the Promotion of Science (JSPS) KAKENHI, Japan (grant number JP26K08290), and the Multidisciplinary Cooperative Research Program (MCRP) in Center for Computational Sciences, University of Tsukuba.

Conflicts of interest:None declared.

**Reference**

1. Fukui, K. *Acc. Chem. Res.* **1981**, *14*, 363–368.
2. Schlegel, H. B. *WIREs Comput. Mol. Sci.* **2011**, *1*, 790–809.
3. Dewyer, A. L., Argüellers, A. J. and Zimmerman, P. M. *WIREs Comput. Mol. Sci.* **2017**, *8*, 1–20.
4. E, W., Ren, W. and Vanden-Eijnden, E. *J. Chem. Phys.* **2007**, *126*, 164103.
5. Behn, A., Zimmerman, P. M., Bell, A. T. and Head-Gordon, M. *J. Chem. Phys.* **2011**, *135*, 224108.
6. Quapp, W. *J. Chem. Phys.* **2005**, *122*, 174106.
7. Henkelman, G., Uberuaga, B. P. and Jónsson, H. *J. Chem. Phys.* **2000**, *113*, 9901–9904.
8. Mitsuta, Y. and Asada, T. *J. Comput. Chem.* **2023**, *44*, 1884–1897.
9. Müller, K. and Brown, L. D. *Theor. Chim. Acta* **1979**, *53*, 75–93.
10. Müller, K. *Angew. Chem. Int. Ed. Engl.* **1980**, *19*, 1–13.
11. Cerjan, C. J. and Miller, W. H. *J. Chem. Phys.* **1981**, *75*, 2800–2806.
12. Simons, J., Jørgensen, P., Taylor, H. and Ozment, J. *J. Chem. Phys.* **1983**, *87*, 2745–2753.
13. Taylor, H. and Simons, J. *J. Phys. Chem.* **1985**, *89*, 684–688.
14. Banerjee, A., Adams, N., Simons, J. and Shepard, R. *J. Phys. Chem.* **1985**, *89*, 52–57.
15. Baker, J. *J. Comput. Chem.* **1986**, *7*, 385–395.
16. Ohno, K. and Maeda, S. Chem. Phys. Lett. 2004, 384, 277–282.
17. Maeda, S., Ohno, K. and Morokuma, K. *Phys. Chem. Chem. Phys.* **2013**, *15*, 3683–3701.
18. Mitsuta, Y., Yamanaka, S., Kawakami, T. and Okumura, M. *J. Comput. Chem.* **2018**, *39*, 1913–1921.
19. Henkelman, G. and Jónsson, H. *J. Chem. Phys.* **1999**, *111*, 7010–7022.
20. Kästner, J. and Sherwood, P. *J. Chem. Phys.* **2008**, *128*, 014106.
21. Quapp, W. A. *Chem. Phys. Lett.* **1996**, *253*, 286–292.
22. Hoffman, D. K., Nord, R. S. and Ruedenberg, K. *Theor. Chim. Acta* **1986**, *69*, 265–279.
23. Jørgensen, P., Jørgensen, H., Jensen, A. and Helgaker, T. *Theor. Chim. Acta* **1988**, *73*, 55–65.
24. Shoji, M., Kayanuma, M. and Shigeta, Y. *Bull. Chem. Soc. Jpn.* **2018**, *91*, 1465–1473.
25. Kikkawa, T., Yamanaka, S., Kawakami, T., Okumura, M. and Shoji, M. *Mol. Phys.* **2026**, in press. https://doi.org/10.1080/00268976.2026.2716330
26. Wang, X., Mao, Y. and Wang, Z. *Chem. Catal.* **2025**, *5*, 101458.
27. Pattanaik, L., Ingraham, J. B., Grambow, C. A. and Green, W. H. *Phys. Chem. Chem. Phys.* **2020**, *22*, 23618–23626.
28. Jackson, R., Zhang, W. and Pearson, J. *Chem. Sci.* **2021**, *12*, 10022–10040.
29. Choi, S. *Nat. Commun.* **2023**, *14*, 1168.

30. Duan, C., Du, Y., Jia, H. and Kulik, H. J. *Nat. Comput. Sci.* **2023**, *3*, 1045–1055.
31. Kim, S., Woo, J. and Kim, W.-Y. *Nat. Commun.* **2024**, *15*, 341.
32. Luo, Y., Gu, X. and Sun, J. *Nat. Commun.* **2026**, *17*, 7597.
33. Hayashi, A., Takamoto, S., Li, J. and Okanohara, D. *J. Chem. Theory Comput.* **2025**, *21*, 1291–1305.
34. Galustian, L., Mark, K., Karwounopulos, J., Kovar, M. P.-P. and Heid, E. *Digit. Discov.* **2025**, *4*, 3492–3501.
35. Xu, L.-C., An, J., Liu, W., Shi, Y.-F., Ji, C.-L., Cao, F. and Qi, Y. *Nat. Commun.* **2026**, *17*, 10254.
36. Yoon, J., Cao, Z., Raju, R. K., Wang, Y., Burnley, R., Gellman, A. J., Farimani, A. B. and Ulissi, Z. W. *Mach. Learn.: Sci. Technol.* **2021**, *2*, 045018.
37. Tang, H., Li, B., Song, Y., Liu, M., Xu, H., Wang, G., Chung, H. and Li, J. *Adv. Sci.* **2024**, *11*, 2304122.
38. Lan, T. and An, Q. *J. Am. Chem. Soc.* **2021**, *143*, 16804–16812.
39. Zhang, J., Lei, Y.-K., Zhang, Z., Han, X., Li, M., Yang, L., Yang, Y. I. and Gao, Y. Q. *Phys. Chem. Chem. Phys.* **2021**, *23*, 6888–6895.
40. Matsumura, Y., Tabata, K. and Komatsuzaki, T. *J. Chem. Theory Comput.* **2025**, *21*, 3523–3535.
41. Xu, W., Zhao, Y., Yan, D., Zhang, X., Chen, J., Wan, Z. and Zhang, R. *J. Comput. Chem.* **2024**, *45*, 487–497.
42. Barrett, R. and Westermayr, J. *J. Phys. Chem. Lett.* **2024**, *15*, 349–356.
43. Kikkawa, T., Kawakami, T., Yamanaka, S. and Okumura, M. *Chem. Lett.* **2025**, *54*, upaf098.
44. Sutton, R. S. and Barto, A. G. *Reinforcement Learning, An Introduction*, 2nd ed.; The MIT Press: London, **2018**.
45. Dong, H., Ding, Z. and Zhang, S., Eds. *Deep Reinforcement Learning, Fundamentals, Research, and Applications*; Springer: Singapore, **2020**.
46. Mills, A. W., Goings, J. J., Beck, D., Yang, C. and Li, X. *J. Chem. Inf. Model.* **2022**, *62*, 3169–3179.
47. Ahuja, K., Green, W. H. and Li, Y.-P. *J. Chem. Theory Comput.* **2021**, *17*, 818–825.
48. Pal, A. *Data Sci.* **2024**, *7*, 73.
49. Mnih, V. et al. *Nature* **2015**, *518*, 529–533.
50. Graves, A. and Schmidhuber, J. *Neural Networks* **2005**, *18*, 602–610.
51. Graves, A., Mohamed, A.-R. and Hinton, G. In *2013 IEEE International Conference on Acoustics, Speech and Signal Processing (ICASSP)*; IEEE: Vancouver, BC, Canada, **2013**; pp 6645–6649.
52. Sutskever, I., Vinyals, O. and Le, Q. In *Proceedings of the 28th International Conference on*

*Neural Information Processing Systems*; **2014**; pp 3104–3112.
53. Karpathy, A., Johnson, J. and Fei-Fei, L. *arXiv:1506.02078* **2015**.
54. Shi, X., Padhi, I. and Knight, K. In *Proceedings of the 2016 Conference on Empirical Methods in Natural Language Processing*; Association for Computational Linguistics: Austin, Texas, **2016**; pp 1526–1534.
55. Adi, Y., Kermany, E., Belinkov, Y., Lavi, O. and Goldberg, Y. *arXiv:1608.04207* **2017**.
56. Moss, G. P. *Pure Appl. Chem.* **1996**, *68*, 2193–2222.
57. IUPAC-IUB Commission on Biochemical Nomenclature. *J. Biol. Chem.* **1970**, *245*, 6489–6497.
58. Yang, Q., Zhang, Y., Dai, W. and Pan, S. J. *Transfer Learning*; Cambridge University Press: Cambridge, **2020**.
59. Lazaric, A. In *Reinforcement Learning*; Wiering, M., Otterlo, M., Eds.; Springer-Verlag: Berlin, Heidelberg, **2012**; pp 143–174.
60. Taylor, M. E. and Stone, P. *J. Mach. Learn. Res.* **2009**, *10*, 1633–1685.
61. Pan, S. J. and Yang, Q. *IEEE Trans. Knowl. Data Eng.* **2010**, *22*, 1345–1359.
62. Zhu, Z., Lin, K., Jain, A. K. and Zhou, J. *IEEE Trans. Pattern Anal. Mach. Intell.* **2023**, *45*, 13344–13362.
63. Taylor, M. E., Jong, N. K. and Stone, P. In *Machine Learning and Knowledge Discovery in Databases*; Daelemans, W., Goethals, B., Morik, K., Eds.; Springer, **2008**; pp 488–505.
64. Kirk, R., Zhang, A., Grefenstette, E. and Rocktäschel, T. *J. Artif. Intell. Res.* **2023**, *76*, 201–264.
65. Garg, S., Bajpai, A. and Mausam. In *Proceedings of the 29th International Conference on Automated Planning and Scheduling (ICAPS)*; **2019**; pp 631–636.
66. Frisch, M. J., Trucks, G. W., Schlegel, H. B., Scuseria, G. E., Robb, M. A., Cheeseman, J. R., Scalmani, G., Barone, V., Mennucci, B., Petersson, G. A., Nakatsuji, H., Caricato, M., Li, X., Hratchian, H. P., Izmaylov, A. F., Bloino, J., Zheng, G., Sonnenberg, J. L., Hada, M., Ehara, M., Toyota, K., Fukuda, R., Hasegawa, J., Ishida, M., Nakajima, T., Honda, Y., Kitao, O., Nakai, H., Vreven, T., Montgomery, J. A., Jr., Peralta, J. E., Ogliaro, F., Bearpark, M., Heyd, J. J., Brothers, E., Kudin, K. N., Staroverov, V. N., Kobayashi, R., Normand, J., Raghavachari, K., Rendell, A., Burant, J. C., Iyengar, S. S., Tomasi, J., Cossi, M., Rega, N., Millam, J. M., Klene, M., Knox, J. E., Cross, J. B., Bakken, V., Adamo, C., Jaramillo, J., Gomperts, R., Stratmann, R. E., Yazyev, O., Austin, A. J., Cammi, R., Pomelli, C., Ochterski, J. W., Martin, R. L., Morokuma, K., Zakrzewski, V. G., Voth, G. A., Salvador, P., Dannenberg, J. J., Dapprich, S., Daniels, A. D., Farkas, Ö., Foresman, J. B., Ortiz, J. V., Cioslowski, J. and Fox, D. J. *Gaussian 09*, Revision C.01; Gaussian, Inc.: Wallingford, CT, **2009**.
67. Rappé, A. K., Casewit, C. J., Colwell, K. S., Goddard, W. A., III and Skiff, W. M. *J. Am. Chem.*

*Soc.* **1992**, *114*, 10024–10035.

68. Cornell, W. D., Cieplak, P., Bayly, C. I., Gould, I. R., Merz, K. M., Ferguson, D. M., Spellmeyer, D. C., Fox, T., Caldwell, J. W. and Kollman, P. A. *J. Am. Chem. Soc.* **1995**, *117*, 5179–5197.
69. Liu, R. and Zhou, X. *J. Phys. Chem.* **1993**, *97*, 1850–1855.

### S-1. Supplementary figures and tables

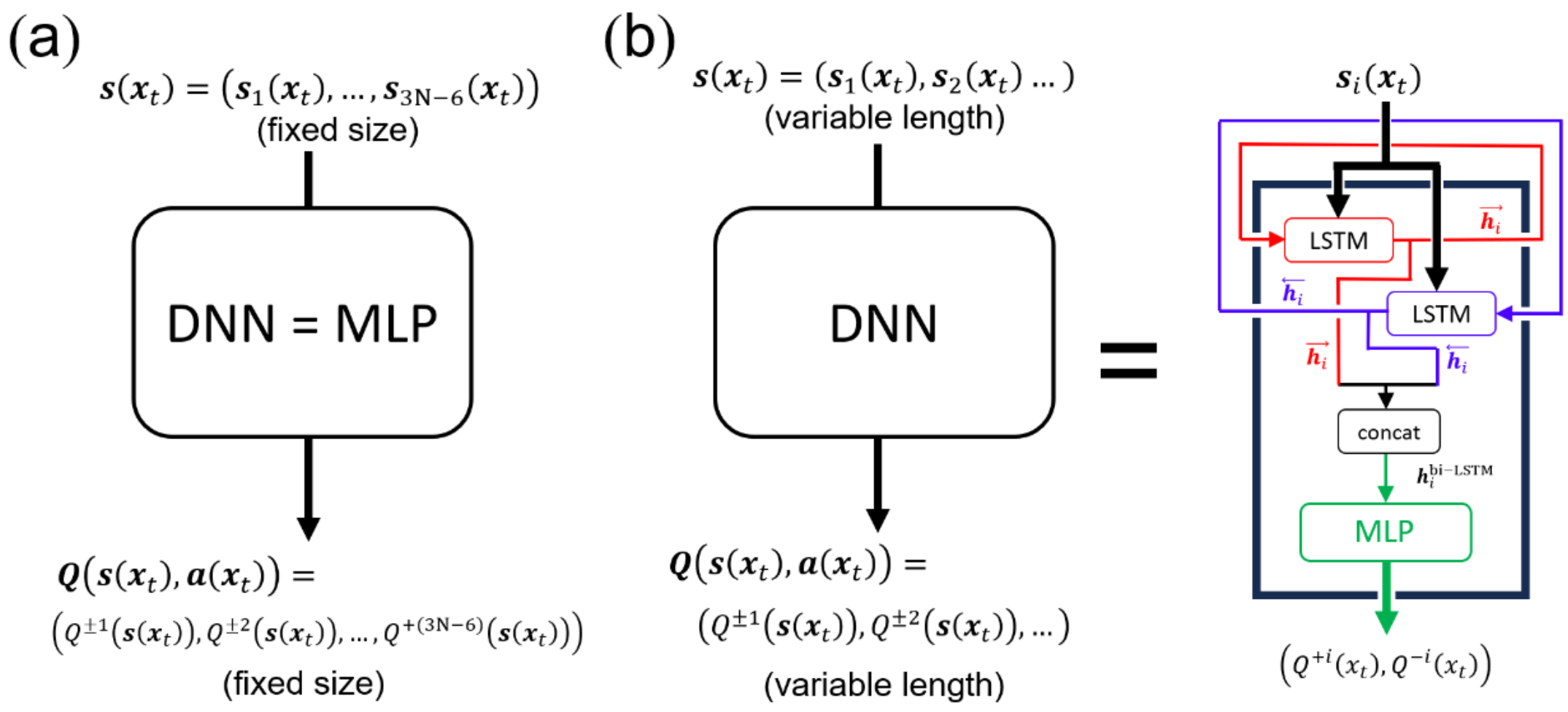

**Figure S1**. Fundamental architectures of our DQN methods for the TS search tasks. (a) Previous model for fixed dimensional problems. (b) Present model for variable dimensional problems. DNN, MLP, and LSTM denote deep neural network, multilayer perceptron, and long short-term memory, respectively.

**Table S1**. Hyperparameters of our model

| Parameter | Value |
|---|---|
| Units per LSTM Layer | 64 |
| Units per MLP Layer | 128 |
| Activation function | tanh |
| Optimizer | Adam |
| Learning rate | $1.0 \times 10^{-3}$ |
| Batch size | 1024 |
| Loss function | Huber Loss |
| Policy update interval | Every 10 steps |
| $\mathbf{Q}^{\mathbf{target}}$ update frequency | Every 1000 steps |
| Replay buffer size | $4.0 \times 10^{4}$ |
| Discount Rate ($\boldsymbol{\gamma}$) | 0.99 |
| $\boldsymbol{\epsilon}$-greedy exploration (without transfer) | $\max\left(0.1, 0.5 - 0.4 \times \frac{t}{2.0 \times 10^{5}}\right)$ |
| $\boldsymbol{\epsilon}$-greedy exploration (with transfer) | 0.1 (constant) |
| Training unit (one round) | 1 episode per equilibrium point (EQ) on the PES |

| Total training rounds | 1000 rounds |
|---|---|

**Table S2**. The coordinates and relative energies for the EQs and TSs of 1,3,5-hexatriene (Triene). Cis (C) and trans (T) conformations are indicated for each EQ.

| | $\Delta E$ (kcal/mol) | $\phi$ (deg) | $\psi$ (deg) |
|---|---|---|---|
| EQ1 (TT) | 0.00 | 180 | 180 |
| EQ2 (CT) | 3.51 | 7.3 | -179.7 |
| EQ3 (CT) | 3.51 | -7.3 | 179.7 |
| EQ4 (TC) | 3.51 | 179.7 | -7.3 |
| EQ5 (TC) | 3.51 | -179.7 | 7.3 |
| EQ6 (CC) | 7.10 | 16.09 | 16.09 |
| EQ7 (CC) | 7.10 | -16.09 | -16.09 |
| TS1 | 9.10 | 89.45 | 179.94 |
| TS2 | 9.10 | 179.94 | 89.45 |
| TS3 | 9.10 | -89.45 | -179.94 |
| TS4 | 9.10 | -179.94 | -89.45 |
| TS5 | 12.52 | 6.6 | 89.81 |
| TS6 | 12.52 | 89.81 | 6.6 |
| TS7 | 12.52 | -6.6 | -89.81 |
| TS8 | 12.52 | -89.81 | -6.6 |

**Table S3**. Trajectory counts reaching each TS region for Triene, categorized by the starting EQ. Asterisks indicate the TS regions connected to each EQ via MEPs, and their total counts are provided in the bottom row. Asterisks indicate the TS regions connected to each EQ via MEPs, and their total percentages are provided in the bottom row.

| | EQ1(TT) | EQ2(CT) | EQ3(CT) | EQ4(TC) | EQ5(TC) | EQ6(CC) | EQ7(CC) |
|---|---|---|---|---|---|---|---|
| TS1 | 279* | 59* | 5 | 3 | 2 | 1 | 3 |
| TS2 | 1084* | 6 | 2 | 3 | 17* | 0 | 3 |
| TS3 | 1137* | 5 | 13* | 12 | 1 | 1 | 0 |
| TS4 | 267* | 2 | 2 | 42* | 9 | 0 | 0 |
| TS5 | 559 | 3854* | 3957* | 13 | 2 | 1000* | 16* |
| TS6 | 186 | 8 | 8 | 1017* | 994* | 3914* | 57* |
| TS7 | 163 | 1007* | 987* | 10 | 9 | 48* | 3903* |
| TS8 | 577 | 7 | 2 | 3858* | 3927* | 15* | 1001* |
| Total for TS regions connected via SDP | 2767 | 4920 | 4957 | 4917 | 4938 | 4977 | 4977 |

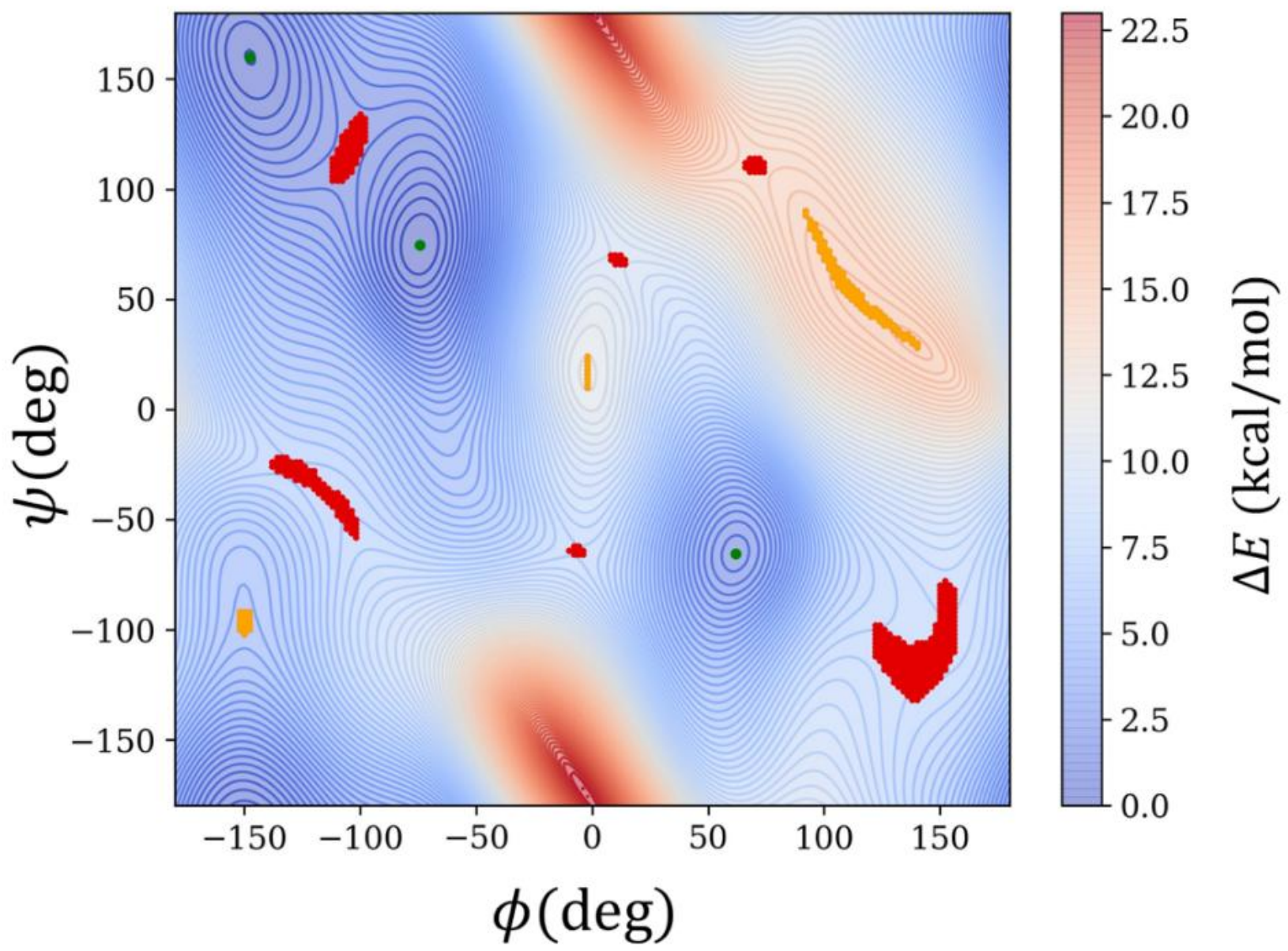

**Figure S2**. Potential energy surface (PES) contour map of alanine dipeptide (Ala2) as a function of the backbone dihedral angles. In addition to the EQ points and TS regions depicted in Figure 6 of the main text, spurious TS regions arising from the loose gradient threshold of our DRL runs are colored in orange (subsequent TS optimizations starting from structures in these regions either converged to other TS regions or failed to converge due to the flatness of the region).

**Table S4**. The coordinates and relative energies for the EQs and TSs of alanine dipeptide (Ala2).

| | $\Delta E$ (kcal/mol) | $\phi$ (deg) | $\psi$ (deg) |
|---|---|---|---|
| EQ1 | 0.00 | -74.36 | 74.52 |
| EQ2 | 0.24 | -147.53 | 159.79 |
| EQ3 | 1.87 | 61.78 | -65.51 |
| TS1 | 1.82 | -105.00 | 117.21 |
| TS2 | 7.03 | -131.09 | -26.92 |
| TS3 | 7.83 | 134.70 | -110.96 |
| TS4 | 7.98 | -6.55 | -64.57 |
| TS5 | 9.78 | 10.62 | 68.12 |
| TS6 | 14.16 | 69.14 | 110.77 |

**Table S5**. Trajectory counts reaching each TS region for Ala2, categorized by the starting EQ (the corresponding counts are listed in Table S3). Asterisks indicate the TS regions connected to each EQ via MEPs, and their total counts are provided in the bottom row.

| | EQ1 | | EQ2 | | EQ3 | |
|---|---|---|---|---|---|---|
| | Non-TL | TL | Non-TL | TL | Non-TL | TL |
| TS1 | 4513* | 4256 * | 4302* | 4788* | 16 | 2 |
| TS2 | 177* | 65 * | 86 | 34 | 53 | 5 |
| TS3 | 12 | 2 | 168* | 115* | 3732 * | 4401 * |
| TS4 | 54* | 112 * | 8 | 3 | 539 * | 431 * |
| TS5 | 116* | 541 * | 6 | 1 | 155 * | 82 * |
| TS6 | 69 | 14 | 296 | 14 | 360 * | 62 * |
| Total for TS regions connected via MEPs | 4860 | 4974 | 4470 | 4903 | 4786 | 4976 |

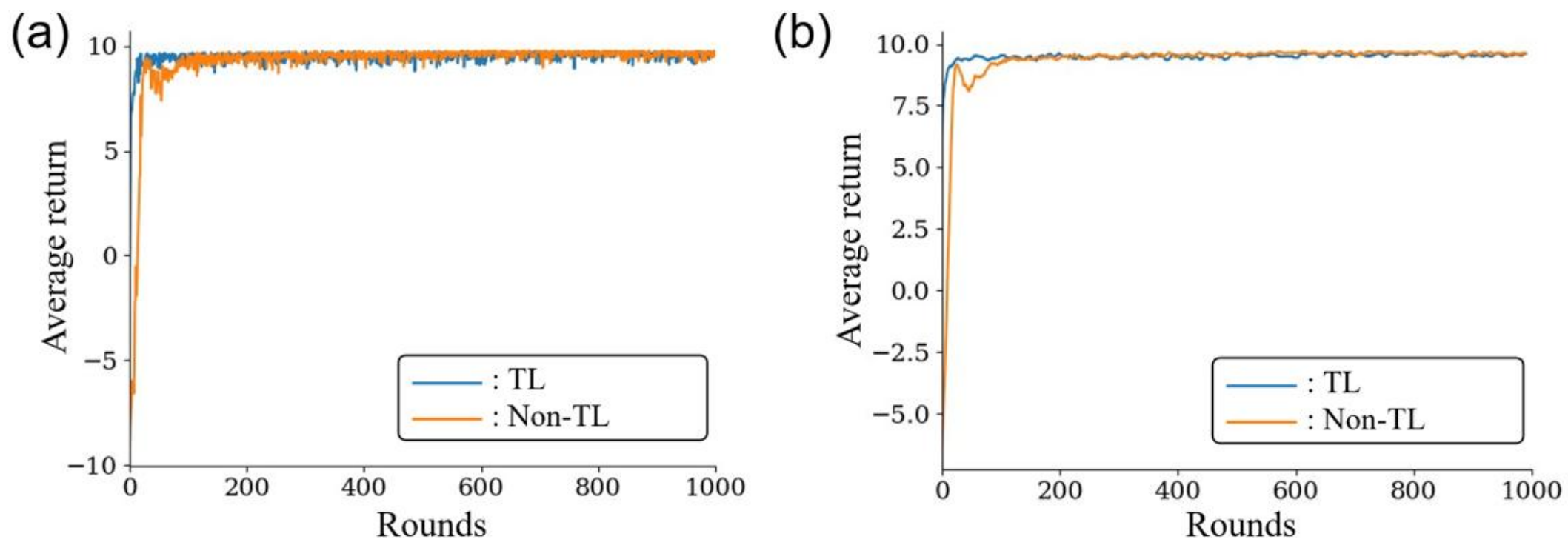

**Figure S3**. Learning curves of the DRL model over the 1000 rounds with transfer-learning (TL) and without it (Non-TL) for 1,3,5,7-octatetraene (Tetraene): (a) average return over 10 independent runs with random initializations and (b) 10-round moving averages of the curves in (a).

**Table S6.** Cartesian coordinates and relative energies for all equilibrium structures (EQs) and transition structures (TSs) of 1,3,5,7-octatetraene (Tetraene). Note: Due to the large volume of data, the complete table is provided as a separate Excel file in the Zenodo repository (https://doi.org/10.5281/zenodo.23726839).

**Table S7**. Number of trajectories and corresponding percentages reaching each TS of 1,3,5,7-octatetraene (Tetraene) for the Non-TL and TL agents. Asterisks indicate the TS regions connected to each EQ via MEPs, and their total percentages are provided in the bottom row. For ease of reference, the complete table is supplied as a separate Excel file in the Zenodo repository (https://doi.org/10.5281/zenodo.23726839)